Larry L Bowman, Jr. and Aimee L Govett
East Tennessee State University

Larry Lee Bowman, Jr., NSF Graduate Fellow in K-12 Education (GK-12 Fellow)
802 Lamont St., Apt 4
Johnson City, TN 37604
bowmanll1@goldmail.etsu.edu

Aimee Lee Govett, Professor
Center for Excellence in Mathematics and Science Education
East Tennessee State University
Box 70684 - Johnson City, TN 37614-1701
govett@etsu.edu

Title: Becoming the Change: A Critical Evaluation of the Changing Face of Life Science, as Reflected in the NGSS

**Abstract**

Tennessee is one of the 26 lead state partners that volunteered to provide leadership and guidance to states for the purpose of adoption of Next Generation Science Standards (NGSS). As stated in the Tennessee Vision for STEM Education (2009), "Tennessee recognizes the importance of science and aims to commit to this understanding by becoming involved in the development, and eventual adoption and implementation of NGSS." The present study correlates the Tennessee State Science Standards to the NGSS for High School Biology/ Life Sciences and examines the need for a dynamic set of standards that teach the technical skills and critical thinking needed in these scientific fields. The NGSS addresses a move from dated scientific quandaries and proposes standards supported by cutting edge scientific research and literature. Partnerships between scientists and educators allow for the information exchange necessary to implement the changes in scientific research in K-12 instruction. Professional development opportunities that include direct partnerships with scientists foster the continued understanding and skills required to teach science.

*Partnerships in K-12 Science Education*

---

Tennessee, as one of the 26 lead state partners that volunteered to provide leadership and guidance to states for the purpose of adoption of Next Generation Science Standards (NGSS), is seriously considering implementing the resulting *NGSS* as presented. The current Tennessee science standards, aligned with (1) National Science Education standards; (2) Benchmarks for Science Literacy, (3) National Association for Educational Progress (NAEP) standards, and (4) the ACT Standards, were adopted in 2007-2008, and implemented during school year 2009-2010. The next implementation cycle of standards will coordinate with the scheduled curriculum materials adoption in school year 2015-2016, which requires the adoption of standards by November 14, 2014. Therefore, Tennessee, similar to many states across the nation, is in the process of reviewing adoption of new science standards according to the NGSS proposed timeline. This illustrates a commitment to adopting a set of standards that can grow and adapt to teaching the skills that scientists need rather than simply supplying factual information. Science changes every day and with it the skills necessary to understand its increasing complexity change. Training the next generation of scientists is not only a daunting task but one that must be abreast of an immense amount of novel research in order to maintain relevance.

This article stems from a grant funded through the National Science Foundation Division of Graduate Education (Grant Number DGE-0742364; P.I. Dr. Gordon Anderson). This NSF GK-12 Graduate Fellowship Program, *Science First!*, is supported by East Tennessee State University in partnership with North Side School of Math, Science, and Technology, a high need and racially/ethnically diverse elementary school. The outcome included GK-12 Fellows in Mathematics, Biology, and Chemistry and K-5 teachers collaborating on developing an integrated curriculum using mathematics and science as a connecting thread (*Access website*: http://www.etsu.edu/cas/gk/).

One of the GK-12 Fellows in Biological Sciences, Larry Bowman (author), has created a series of posters showing correlations between the current Tennessee Science Education standards (Huffman, 2009) and NGSS (Achieve, 2013). The sequence of 18 posters serves as a series of guidemaps between the Tennessee Curriculum Standards for Science Education (TNCSSE) for grade levels kindergarten through high school and the corresponding NGSS (See http://www.netstemhub.com/). The 18 guidemaps were presented at the recent Tennessee Science Teachers Association (TSTA) conference (November 2013). During the November TSTA conference, Dr. Scott Eddins, State Board of Education (SBE) staff member, discussed the SBE Science Education Review plan. The steering committee, appointed by SBE, has asked to use the collection of guidemaps in their deliberation of the science standards review, adoption, and implementation.

The TNCSSE-NGSS Guidemaps were also presented at a one-day regional symposium for teachers, administrators, and other interested parties (December 2013) to generate dialogue at the local level of implementation issues for both Common Core State Standards (CCSS) and NGSS. The breakout sessions were divided among Secondary 7-12 Math, Middle and Secondary Science, and Elementary STEM/Language Arts. In addition to the wide-range contention of the need for intensive professional development focused on NGSS, the notes from the Middle and Secondary Science breakout group are summarized in the following table:

INSERT TABLE 1

*Professional Development for Teachers*

In this article, we will focus on the guidemap showing the correlation of the TNCSSE for Biology I (Huffman, 2009) to the NGSS for High School Biology/Life Sciences (Achieve, 2013) and examine the need for a dynamic set of standards that teach the technical skills and critical thinking needed in these scientific fields. The NGSS addresses a move from dated scientific quandaries and proposes standards supported by cutting edge scientific research and literature (Bybee, 2011; Bybee, 2012; Bybee, 2013a; Willard, 2013). Partnerships between scientists and educators allow for the information exchange necessary to implement the changes in scientific research in K-12 instruction. Professional development opportunities that include direct partnerships with scientists foster the continued understanding and skills required to teach science (Zhang, McInerney, & Frechtling, 2010).

New curriculum standards are generally communicated through a top-down approach. Consequently, there is a need for continuing support for local administrators and teachers, who are responsible for carrying out large-scale educational change classroom by classroom, in the form of a professional learning community so that the endorsement and understanding of the new ideas such as NGSS do not become distorted or misinterpreted. The creation of a functional professional learning community has the explicit intent of providing opportunities for teacher learning. Richmond and Manokore (2011) established essential components for functional and sustainable professional learning communities, from which we will focus on teacher learning and collaboration. Teachers must be partners in their own professional development and the shared goal should be to increase their knowledge and improve their self-efficacy. They need to be confident of their ability to organize and clarify to their students the often ambiguous and complex ever-changing nature of scientific practices today (McNeill & Knight, 2013).

Science teachers are charged with keeping education up to date with a discipline that is always changing and staying abreast of the changes in a specific discipline. Experienced teachers can craft their own curriculum based on state standards given an ideal teaching environment. Teachers, as curriculum and pedagogical experts, address all new state prescribed criteria with the dichotomy of a desire to teach what is current and right but also not to teach what is incorrect, or not grade level appropriate. As Rodger Bybee (2013a), NGSS Writing Team Co-Leader for Life Sciences, emphasizes in a recent article in Science and Children, teachers need a viable curriculum in order to implement the standards properly and with confidence.

## *TNCSSE-NGSS Guidemaps*

The TNCSSE-NGSS Guidemaps are designed with ease of use in mind. Care was taken to ensure that the guidemaps are a more effective way to correlate standards than perusing through either state (Huffman, 2009) or NGSS's websites (Achieve, 2013). Educators can use the guidemaps as a way to adapt their current and past lesson plans to NGSS standards by easily visualizing inclusions and exclusions between the two sets.

The TNCSSE standards are pictured in center left (Figures 1.1, 1.2, 1.3, 2, and 3) by Course Level Expectation (CLE) with their corresponding explanations to the far left (Huffman, 2009). The NGSS standards are pictured in center right (Figures 1.1, 1.2, 1.3, 2, and 3) by Disciplinary Core Idea (DCI) with their corresponding explanations to the far right (Achieve, 2013). In the center of the guidemap, the interaction between the TNCSSE and NGSS can be seen. If an arrow points from a TNCSSE CLE to a NGSS DCI, the two correlate on some level. If there is no arrow protruding from a TNCSSE CLE, the standard was excluded from the NGSS. Conversely, if there is no arrow intercepting an NGSS DCI, the standard is an entirely new inclusion within the NGSS. Because the standards systems are not directly correlative, several have multiple arrows feeding out of and into others. Areas where this occurs have high "coverage" meaning that the knowledge implied is thoroughly addressed in both sets of standards. Likewise, a caveat of the maps is that though there is some coverage, every arrow is not equally weighted, e.g., a TNCSSE CLE may correspond to only a small portion of a much more detailed NGSS DCI. A close reading of the coverage gives a more thorough understanding or the similarities and difference between the two standards sets as further discussed here within.

INSERT FIGURE 1.1

## *Changes in High School Biology Curriculum Standards*

### *Changes in Cellular and Molecular Biology*

The first standard to be excluded by the NGSS is CLE 3210.1.1 (Table 1.1), which states that students should be able to "compare the structure and function of cellular organelles in both prokaryotic and eukaryotic cells" (Huffman, 2009). Though this distinction between cellular types has been understood and taught in this manner for decades, it was proposed in 1990 that there are three distinct domains of life through molecular resolution, specifically ribosomal RNA (rRNA) analyses (Woese, Kandler, & Wheelis, 1990). The understanding of organisms as either prokaryotic or eukaryotic involves the separation of life based on a single phenotypic characteristic whereas, the three-domain system further resolves organisms previously classified as prokaryotes into two distinct groups via molecular analyses. Though the differences between prokaryotic and eukaryotic cells is not at odds with the three-domain classification system, the focus has changed from using a single, phenotypic trait to using molecular evidence to classify organisms.

Our current understanding of the classification of life involves similarities and differences of structure and function of three distinct life domains, rather than a two-domain system, and thus, this standard's exclusion by the NGSS is justified as a dated understanding of classification systems (Figure 1.1). This shift in the NGSS reflects a shift in the life sciences to include molecular approaches in addition to morphology, behavior, and physiology to resolve unanswered questions in the tree of life and evolutionary biology questions. The standard in the TNCSSE addresses the difference between cellular organelles and functions between two cell types, as part of a dichotomous two-domain system. However, scientists have recognized a three-domain system, which includes Bacteria, Archaea, and Eukaryota, since the early 1990s (Sapp & Fox, 2013), and teaching a three-domain system is imperative to understanding the current classification of life. We see in the NGSS that distinguishing between prokaryotes and eukaryotes is useful and necessary but not as the basis for a dichotomous classification system. Also noted in the NGSS is the need to know and understand the function and structure of organelles regardless of domain.

The exclusion of this topic, however, still raises questions about the methodology of teaching it effectively. A proposed method of teaching the topic does not vary much from the present model except for one significant element: the focus is on organelle structure and function and not on using organelle phenotypes to resolve organisms into different groups. It is equally important to know that the three-domain system exists and how it is resolved, yet for the purposes of high school biology, understanding the molecular basis for the distinction may not be developmentally appropriate. Moreover, the specifics of molecular genetics' use as a tool for classification is forever advancing and has now spawned its own disciplines, such as transcriptomics, lipidomics, and metabolomics. This is an example where the NGSS reflects recent research and ideas that are progressing in the scientific community, while maintaining relevance and age-appropriateness for students.

INSERT FIGURE 1.2

Other inclusions to the NGSS include our current broader understanding of molecular and cellular topics including: macromolecules, cell development, and enzymes. Specifically the NGSS calls for a deeper focus on understanding how all things are made of the same organic parts rearranged in different ways (Figure 1.1) (Achieve, 2013). The focus is not only on composition but how molecules interact and change into other macromolecules (HS-LS1-2 and HS-LS1-6). The previous understanding taught at the high school level was the identification of the four major macromolecules: proteins, carbohydrates, nucleic acids, and lipids (CLE 3120.1.2) (Huffman, 2009). The deeper understanding that the NGSS encourages directly reflects current research in molecular mechanics and understanding chemical reactions and interactions emerging in the field of biophysics and others (Alberts, 1998; Mattick, 2007). The NGSS additionally encourages a change in the instruction of cell development. Our ability to detail gene expression at a singular cell level will only continue to increase our understanding of cellular development (Elowitz, Levine, Siggia, & Swain, 2002; Shapiro, Biezuner, & Linnarsson, 2013). Our new understanding of epigenetics and translational control of real-time cell development at high resolution (Pazdernik & Schedl, 2013) requires the need to teach not only the steps of cell division and development but also the hierarchal effects that every cell has on the overall organism (HS-LS1-1-4) (Achieve, 2013).

A great breadth of research released from the Encyclopedia of DNA Elements project (Feingold et al., 2004) has allowed for the further emphasis on the regulatory functions of RNA (Mattick, 2009). Several studies have reported transcriptional and translational control mechanisms on cellular processes that could be more important than enzymatic controls for some cellular processes (Hansen et al., 2013; Neph et al., 2012; Pelechano, Wei, & Steinmetz, 2013). Moreover, cellular products previously thought to be "dead enzymes," i.e., those with enzymatic structure but lacking binding sites or other crucial structures for function, now are understood to have definite roles in chemical pathways, mostly regulatory in nature (Leslie, 2005). Thus, there is a need to teach the processes of homeostasis (HS-LS1-1-6) in broader and more open-minded ways that include the ability to propose new mechanisms and interactions between molecules that were previously thought to have little effect on molecular regulation. The NGSS encourages the use of modeling and inquiry-based learning that the next generation of scientists will need to continue unlocking the secrets of cellular development and molecular regulation (Figure 1.1) (Achieve, 2013). The NGSS is a dynamic set of standards that can and will adapt to the changing face of science especially with its focus on proposing novel models to explain biological phenomena. Learning skills, such as technical reading, interpretation, critical thinking, and analysis rather than factual learning is only further emphasized by the wildly increasing rates at which "facts" are changing.

INSERT FIGURE 1.3

### *Changes in Genetics and Inheritance*

The changes in our understanding of genetics and patterns of inheritance in the last decade are so vast that resolving them is even more challenging. How to teach a paradigm-shifting multitude of new information to our students will seem even more unrealistic. However, with the correct tools and motivations teaching the new innovations in genetics will assuredly be invigorating. There have been previous, extensive studies on how to best teach the ever elusive topic of genetics (Duncan & Tseng, 2011), however, now much of the information we once thought was true is now considered debatable and even questionable (Portin, 2009). Again, with the continued release of the wealth of information from the ENCODE project results; we are in the midst of a paradigm shift in the genetics world (Birney et al., 2007; Feingold et al., 2004). Despite resistance by some scientists against the newly released information (Doolittle, 2013; Eddy, 2013; Graur et al., 2013), a scientific revolution is underway (Kuhn, 1970).

The Central Dogma (DNA transcribed into mRNA translated into protein) proposed by Francis Crick in 1970 is no longer the *status quo* (Crick, 1970; Mattick, 2009), yet it is still the predominantly taught model for genetics (CLE 3210.4.1-2) (Huffman, 2009). Though the science behind the Central Dogma is sound and works for many systems, it is slowly becoming the accepted standard that it is merely one piece of a larger puzzle and may, in fact, be a less important pathway than other regulation pathways. Our current understanding of genes is that they are merely parts of vast gene regulatory networks interacting together to produce the molecular products necessary for metabolism (Davidson & Erwin, 2006). We also have a much better understanding of as large portion of the genome that was previously classified as "junk

DNA" (Ohno, 1972) despite its highly conserved nature (Bejerano et al., 2004), that has now been proposed to have far-reaching implications as regulatory mechanisms in gene regulatory networks (Djebali et al., 2012; Kolata, 2012; Portin, 2009). Transcripts and transcriptional regulation are also emerging fields of their own that have ramifications for inheritance and gene expression. The most current research on transcripts and transcript isoforms posits them as important regulatory mechanisms that are active in molecular regulation, epigenetic expression, and inheritance probabilities (Djebali et al., 2012; Pelechano et al., 2013). The increasing importance of RNA in biological systems continues to be elucidated (Hansen et al., 2013) in a field that was largely anchored by the perceived importance of DNA. New next generation sequencing techniques and other innovative molecular techniques have allowed for the emergence of comparative genomics as a field (Metcalfe, Filée, Germon, Joss, & Casane, 2012).

INSERT FIGURE 2

Contemporary genetics is a more complex and perpetually changing field than it was even five years ago. Deciding what to teach young scientists is becoming increasingly difficult as the amount of information quickly outpaces the ability for such information to be updated in textbooks and in classrooms. In the NGSS, however, we see a set of genetics and heredity standards that equally drop old understandings of genetics and propose new standards with loose boundaries that are easily capable of expanding to the additional information as it becomes available (HS-LS3-1-3, Figure 2) (Achieve, 2013). We see a distinct movement away from the very rigid models of inheritance proposed by the Punnett square and Mendelian genetics (CLE 3210.4.3-4) and the Central Dogma (CLE 3210.4.1-2, Figure 2). Instead of proposing rigid inheritance schemes, the NGSS focus on modeling and clarifying the fact that many schemes are not only possible but likely (HS-LS3-1-3, Figure 2) (Achieve, 2013; Huffman, 2009). There is a change from a parent-offspring level to observing and predicting how populations grow and adapt, using statistics and probability, which is a common method in scientific research and a much more practical skill in the work force, too. Noted also are the emphases on epigenetics and defending proposed models (HS-LS3-2) (Achieve, 2013). With an infinite amount of possibilities for genetic expression and inheritance seemingly possible, the focus now shifts from possibility to probability and defense of that probability through evidence-based claims and statistics.

In addition to the important changes and updates the NGSS fosters, we do see a noticeable exclusion from the TNCSSE: "Assess the scientific and ethical ramifications of emerging genetic technologies" (CLE 3210.4.7, Figure 2, Table 1.1) (Huffman, 2009). This exclusion, however, is both a removal of dated information and a developmentally inappropriate topic. Genetic engineering technologies and genetically modified organisms is no longer a rare topic but a commonplace occurrence. We now know the scientific ramifications of genetic sequencing and the mysteries it has unlocked for scientists. Moreover, the ethical ramifications of genetically modified organisms are a highly debated subject and one that scientists must engage with and police. The debate has changed from when genetic engineering will become common to how far should genetic engineering be allowed to progress and remain moral; an ethical morality debate by high school biology students is not a testable standard. High school freshman are not developmentally capable of debating questions of morality with hopes of concrete resolutions. We recognize the need to foster scientific argumentation and meaningful debate among young scientists. However, students do not have the wherewithal or understanding of the intricacies involved in current genetic engineering procedures or outcomes to make rational, moral judgments on its use as a scientific method, justifying this moral debate's exclusion from the high school curriculum. Should genetic engineering happen is no longer a question because it has and continues to occur. But, genetic engineering's current repercussions are not developmentally appropriate for high school biology students to resolve.

The changes in teaching methodology for these subjects will take time and patience. Genetics, more than any other, is progressing at a rate previously unperceiveable, due to incredible advances in genetic sequencing technologies. Keeping teachers ahead of this knowledge will be difficult, which is why we believe the NGSS downsizes genetics and inheritance into digestible standards and leaves more advanced understandings of genetics networks for undergraduate preparation. Specifically, the NGSS does not require the teaching of specific and rigid inheritance schemes but emphasizes the need for new inheritance models that can adapt and change as new information becomes available (Achieve, 2013). The NGSS promote student-generated models with evidence-based explanations, which is the same model that geneticists use to justify their own findings. As it seems, with our quickly advancing understanding of inheritance and gene expression, genetics is not a cookie-cutter field with overarching mechanisms but a dynamic one filled with intricate and various methods to accomplish similar goals across the tree of life. Thus, a set of rigid cookie-cutter standards will no longer effectively teach such a discipline. The resulting NGSS leaves room to grow and adapt as our understanding of genetics continues to increase in complexity.

### *Changes in Ecology, Evolution, and Biodiversity*

Recent climate change events make changes in ecology, evolution, and biodiversity findings and their subsequent standards changes especially relevant. A major inclusion with the NGSS is to "evaluate the evidence for the role of group behavior on individual and species' chances to survive and reproduce" (HS-LS2-8, Figure 1.3, Table 1.1) (Achieve, 2013). Animal behavior is at the crossroads of physiology and environment, and with current research on epigenetic effects as mentioned previously, how animals react to their environments individually and as communities and populations can greatly affect their ability to survive great climatic shifts (Colbourne et al., 2011; Doorslaer, Stoks, Duvivier, Bednarska, & De Meester, 2009; Pinsky, Worm, Fogarty, Sarmiento, & Levin, 2013). This inclusion is important to understand the current state of many biodiversity hotspots on the planet that are incurring major climatic shifts (Hampton et al., 2008; Kozhov, Kozhova, Izmest'eva, & Izmest′eva, 1998; Matzinger, Spirkovski, Patceva, & Wüest, 2006; McKinnon, 2002; Moore et al., 2009; Pinsky et al., 2013; Shimaraev, Kuimova, Sinyukovich, & Tsekhanovskii, 2002; Tierney et al., 2010; Verburg, Hecky, & Kling, 2003). The ability to predict the subsequent responses is an unavoidable consequence of anthropogenic climate change, and as an inclusion of this standard in the NGSS speaks directly to the need to develop the young scientists' ability to think about questions such as biodiversity and survivability.

INSERT FIGURE 3

Additional changes to the ecology, evolution, and biodiversity curriculum standards include the exclusion of "explore the evolutionary basis of modern classification systems" (CLE 3210.5.6, Figure 3, Table 1.1) (Huffman, 2009) and the inclusion of "create or revise a simulation to test a solution to mitigate adverse impacts of human activity on biodiversity" (HS-LS4-6, Figure 3, Table 1.1) (Achieve, 2013). The exclusion of CLE 3210.5.6 is linked to advances in genetic sequencing in the aforementioned changes in genetics and inheritance. With new advances in our ability to sequence entire genomes, it is becoming increasingly easier to systematically classify organisms on a molecular level instead of a phenotypic level alone (Lemmon & Lemmon, 2013; Yang & Rannala, 2012). The continually developing discipline of phylogenetics deals with precisely these consequences of innovative sequencing techniques (Wiley & Lieberman, 2011). However, along with the complexity of contemporary genetics, it is beyond the scope of high school biology to study comparative genomics, and thus, there is not an equivalent, updated phylogenetics standard in the NGSS. However, there is an updated standard that directly addresses anthropogenic climate change: the inclusion of HS-LS4-6, which requires students to not only understand human impact on the environment, but also to test solutions via simulation modeling (Achieve, 2013). This is an extremely practical and relevant skill in the scientific community, and the instruction of simulation modeling to solve problems has myriad advantages in the non-scientific job market, as well.

    The final inclusion of the NGSS in this realm is the specific and finite instruction of evidence-based explanations for evolution (HS-LS4-2, Figure 3, Table 1.1) (Achieve, 2013). The TNCSSE calls for a "summar[y of] the supporting evidence for the theory of evolution" (CLE 3210.5.4, Figure 3, Table 1.1) (Huffman, 2009) but neglects to outline the four factors that can lead directly to evolution as the NGSS standard does. In the scientific community, the amount of evidence-based examples and explanations for evolution abound, so the focus has changed from supporting evolution as a theory to supporting how evolution occurs. The NGSS assumes an acceptance of evolution and further encourages explanations that do not seek to prove but seek to understand. The understanding of life history and specific strategies and costs that species' undergo when under selection is an invaluable skill to learn in high school biology

and is at the core of understanding every other discipline of biology from cellular and molecular to ecosystem dynamics. The critical thinking alone that this standard requires of students is enough to support its inclusion in the NGSS, but as an essential building block for understanding higher level biology, its exclusion would be a travesty and hinder our students' success in higher education.

The proposed method of teaching the changes to the ecology, evolution, and biodiversity curriculum will be much more in-depth than the previous TNCSSE. Where a lot of detail was eliminated in the genetics and inheritance curriculum, an equal amount was added to the ecology, evolution, and biodiversity curriculum. The NGSS goes far deeper into the understanding of how organisms interact with the living and nonliving components of their environments. Teaching "big picture" science often is overshadowed by the "microsciences" in the high school curriculum where emphasis is commonly on cellular and molecular level biology. However, the NGSS fills out ecology well and the need to produce "big picture" scientists for tomorrow. As questions of our own future are raised in response to the rapidly declining biodiversity of biomes all over the planet, the inclusion of new standards and increased depth is a welcome change.

### *Changes in Scientific and Intellectual Inquiry*

The NGSS does not supply specific inquiry standards as the TNCSSE standards did in the past. This may, at first, present a glaring problem for many educators and policymakers looking to adopt a new set of standards for teaching science. However, after analyzing the NGSS extensively, one finds that there is more inquiry embedded in the standards than was previously explicitly stated by the TNCSSE. The inquiry of NGSS is embedded in such a way that its instruction is coupled with the correct instruction of other standards. In the TNCSSE, with explicit and separate inquiry standards, the challenge was "how [to] get teachers to think of content and inquiry as not mutually exclusive, but rather aspects of the same goal" (Quigley, Marshall, Deaton, Cook, & Padilla, 2011). Since the NGSS *is* inquiry-based, simply teaching the standards as written requires an embedded inquiry that surpasses the TNCSSE explicit standards.

    Specific examples from high school biology and aforementioned include: HS-LS1-3 (Figure 1.1), HS-LS3-2 (Figure 2), and HS-LS4-2 (Figure 3, Table 1.1) (Achieve, 2013). In HS-LS1-3, students are required to understand homeostasis through investigations and feedback mechanisms of macromolecules. Not only are students required to understand the mechanics of cellular homeostasis but also how those mechanics interact to cause the chemical cascades necessary to begin and maintain homeostasis. In HS-LS3-2, students are required to propose their own models, such as epigenetics, for genetic inheritance and variation and then to demonstrate how different effects might change the outcome. Students are learning contemporary genetics understanding in a deeply critical mindset, having to propose theoretical models for witnessed phenomena. In HS-LS4-2, students are expected to support an explanation for speciation and selection using one or more of the four examples given which requires students to not only understand each factor that can effect evolution but also how those factors interact synergistically to enhance or hinder the process.

    With NGSS, these taken as only select examples among many, there is no longer a product-based learning environment but inquiry-based endeavors that require students to ask their own questions which are supported by observation and investigate the responses which are supported by their observations. This cyclical method of learning is exactly the methodology used by academic and industrial scientists and implementing this type of organic inquiry in the K-12 curriculum especially in high school biology is not only a necessary skill but also will

encourage student ownership of the subject matter (Wyner, 2013). Increasing rigor in the classroom, especially in the sciences, has been shown to increase the quality of students' scientific argumentation (Sampson, Grooms, & Walker, 2011), which is evidence of a general ability to think logically.

As mentioned, this cyclical method of inquiry-based learning prepares students for various fields in academic and industrial science. Inquiry-based learning is also a model of how to learn scientific vocabulary with regular practice with complex texts and its academic language. When a research scientist asks a question, he or she begins with a question and then asks if anyone has previously answered that exact question. If so, how did they describe it and what words did they use. Those words then become the scientific vocabulary for describing that particular phenomenon. If not previously answered, he or she designs an experiment to test the question but still looks to the literature to obtain ideas about how others have approached the problem and how they described the phenomenon, again adopting the science vocabulary. Students demonstrate a more detailed knowledge base of the material when they are subject to experiential inquiry-based learning (Nadelson, 2009).

This proposed method of teaching the NGSS requires students to come to an organic understanding of the science and the science vocabulary they are learning because they have an invested interest in answering their own questions, which the NGSS encourages them to ask. A particular challenge for scientific vocabulary is the propensity for scientists to engage different words to describe similar or exact things. Bybee addresses the subtlety of changing "*abiotic* and *biotic*" to "*nonliving* and *living*" and how that simple change, though seemingly trivial, helps align meaning with terminology: "by aligning the words, we align the meaning, and, in the end, the student understandings intended by the NGSS" (Bybee, 2013b). The NGSS will teach students to not be timid of science vocabulary but to embrace it as a tool to understand and precisely discuss the phenomena that they witness in the natural world around them. Students will recall vocabulary more effectively because they specifically had to search for words to describe what they wanted to say, organically developing their own scientific vocabulary (Pease & Kuhn, 2011). Further, the critical thinking that the NGSS fosters is a model for how real science is conceived and executed. The implementation of the NGSS would better prepare students for the rigors of undergraduate science classes which require them to not only recall but apply their knowledge.

*Discussion/Conclusion*

For STEM faculty in higher education institutions, tenure and promotion policies tend to weigh heavily on research, with service to the public, in this case K-12 schools, a very distant last (Zhang et al., 2010). Scientists and those in the professional development community need to change their habits of mind as well. What the scientific community, e.g., American Association for the Advancement of Science (AAAS), National Science Foundation (NSF), National Center for Improving Science Education (NCISE), and others, desires to foster is the growth of the next generation of scientists. In general, connecting scientific research to undergrad education is becoming an essential component of research grants. Several directives at NSF also focus on undergrad STEM education, such as the Improving Undergraduate STEM Education (IUSE) program. Emphasis on inquiry and critical thinking cannot be successful if it starts in college.

Therefore, forming partnerships that involve K-12 districts, teachers and administrators, as well as STEM and Education faculty in institutions of higher education are essential (Richmond & Manokore, 2011). The importance of communication and coordination with K-12 educators may augment scientists' views of teachers as professionals and deepen scientists' pedagogical orientations (Schuster & Carlsen, 2009). It is in college faculties' best interest to interact with K-12 teachers to improve the preparedness of their own incoming freshman. The challenge is how to bring those very different habits of mind together to bring about a sustainable professional learning community (Supovitz & Turner, 2000).

Becoming the Change:  23
A Critical Evaluation of the Changing Face of Life Science, as Reflected in the NGSSKuhn, T. S. (1970). The structure of scientific revolutions. *Chicago/London.*

Lemmon, E. M., & Lemmon, A. R. (2013). High-throughput genomic data in systematics and phylogenetics. *Annual Review of Ecology, Evolution, and Systematics, 44*(1).

Leslie, M. (2005). New trick for an old enzyme. *Science's SAGE KE, 2005*(22), nf41.

Mattick, J. S. (2007). A new paradigm for developmental biology. *Journal of Experimental Biology, 210*(9), 1526-1547.

Mattick, J. S. (2009). Deconstructing the dogma. *Annals of the New York Academy of Sciences, 1178*(1), 29-46.

Matzinger, A., Spirkovski, Z., Patceva, S., & Wüest, A. (2006). Sensitivity of ancient lake ohrid to local anthropogenic impacts and global warming. *Journal of Great Lakes Research, 32*(1), 158-179.

McKinnon, J. S. (2002). Aquatic hotspots: Speciation in ancient lakes III. *Trends in Ecology & Evolution, 17*(12), 542-543. doi:http://dx.doi.org/10.1016/S0169-5347(02)02643-5.

McNeill, K. L., & Knight, A. M. (2013). Teachers' pedagogical content knowledge of scientific argumentation: The impact of professional development on K–12 teachers. *Science Education, 97*(6), 936-972.

Metcalfe, C. J., Filée, J., Germon, I., Joss, J., & Casane, D. (2012). Evolution of the australian lungfish (neoceratodus forsteri) genome: A major role for CR1 and L2 LINE elements. *Molecular Biology and Evolution, 29*(11), 3529-3539.

**Table 1:** Questions and concerns from the Middle and Secondary Science breakout sessions at a regional symposium on implementation issues for both Common Core State Standards and Next Generation Science Standards (December 2013).

| Concerns regarding NGSS | Questions about NGSS |
|---|---|
| Need administrative support | How to provide basic knowledge? |
| Need autonomy to teach science using evidence; finding and using information. | How not to leave students with misconceptions? |
| Time management: more depth less breadth | How do we survive (drill & kill / to test; keep our jobs) while teaching <u>real</u> science thinking? |
| Need differentiation - not all HS degrees are the same. Assessments don't fit (work for) all students and do not represent career options. | How will students be tested? (Portfolio assessments, presentations, model building…) |
| Retention vs. Social Promotion | Can math and problem solving be taught or is innate aptitude a constraint? To what extent? |
| Need for <u>college level and vocational</u> courses in high school! Need ongoing provision of necessary equipment and technology. | How do we get family and community accountability not just teacher (accountability)? |

**Table 2:** High School Biology curriculum changes with the adoption of Next Generation Science Standards (NGSS), Course Level Expectations (CLE) dropped by the NGSS (exclusions) and Disciplinary Core Ideas (DCI) added (inclusions).

| Exclusions | Inclusions |
|---|---|
| *CLE 3210.1.1* Compare the structure and function of cellular organelles in both prokaryotic and eukaryotic cells. | *HS-LS1-2* Develop and use a model to illustrate the hierarchical organization of interacting systems that provide specific functions within multicellular organisms. |
| *CLE 3210.4.7* Assess the scientific and ethical ramifications of emerging genetic technologies. | *HS-LS2-8* Evaluate the evidence for the role of group behavior on individual and species' chances to survive and reproduce. |
| *CLE 3210.5.5* Explore the evolutionary basis of modern classification systems. | *HS-LS4-2\** Construct an explanation based on evidence that the process of evolution primarily results from four factors: (1) the potential for a species to increase in number, (2) the heritable genetic variation of individuals in a species due to mutation and sexual reproduction, (3) competition for limited resources, and (4) the proliferation of those organisms that are better able to survive and reproduce in the environment. |
|  | *HS-LS4-6* Create or revise a simulation to test a solution to mitigate adverse impacts of human activity on biodiversity. |

*Standard previously listed in National Science Education Standards (National Research Council, 1996) but not in Tennessee Department of Education Curriculum Standards in Science Education (Huffman, 2009).

**Figure 1.1**

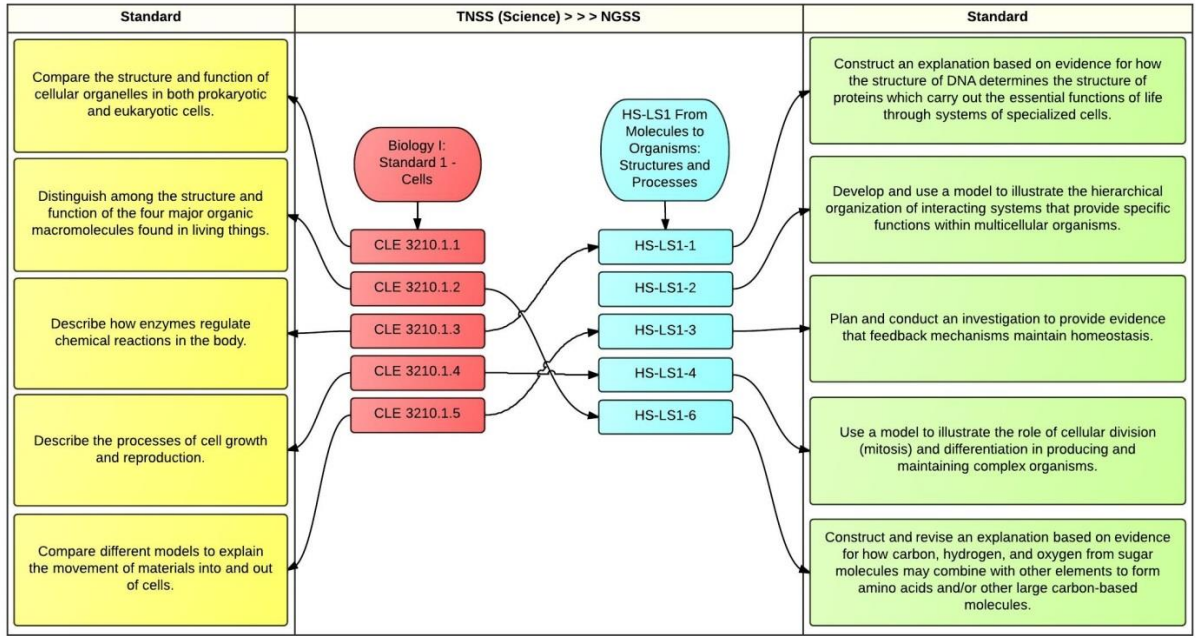

**Figure 1.1:** High School Biology Guidemap for Cellular and Molecular Biology (Part 1), Tennessee Department of Education Curriculum Standards in Science Education (left center) with explanations (far left) correspondence to Next Generation Science Standards (right center) with explanations (far right).

**Figure 1.2**

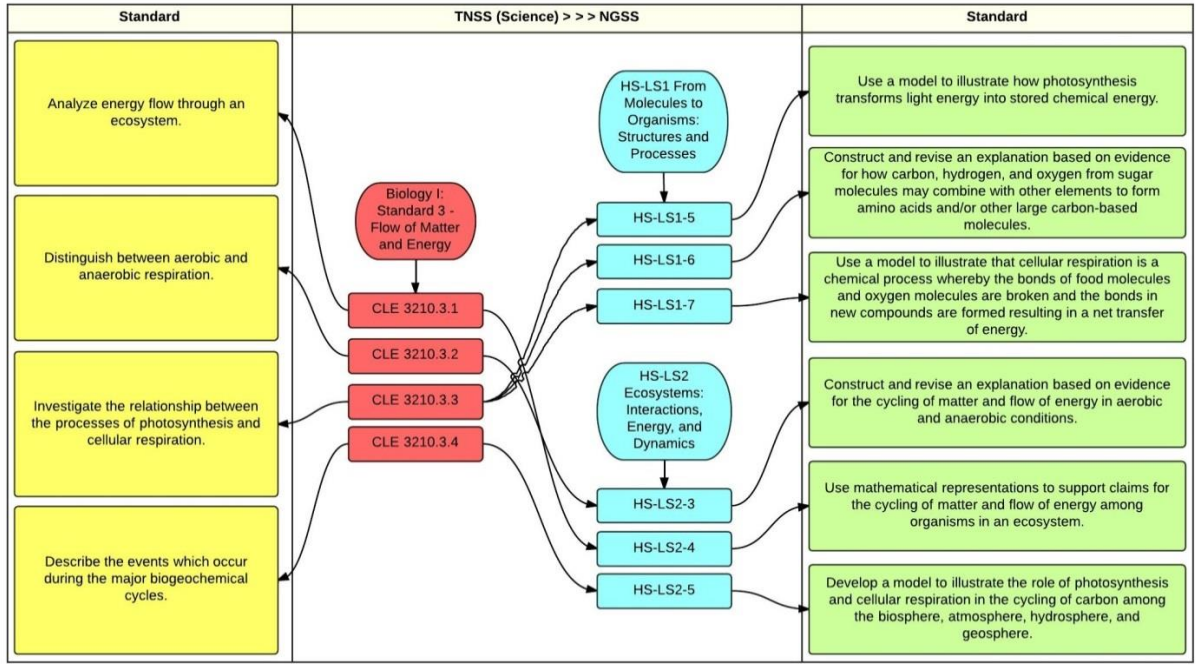

**Figure 1.2:** High School Biology Guidemap for Cellular and Molecular Biology (Part 2), Tennessee Department of Education Curriculum Standards in Science Education (left center) with explanations (far left) correspondence to Next Generation Science Standards (right center) with explanations (far right).

**Figure 1.3**

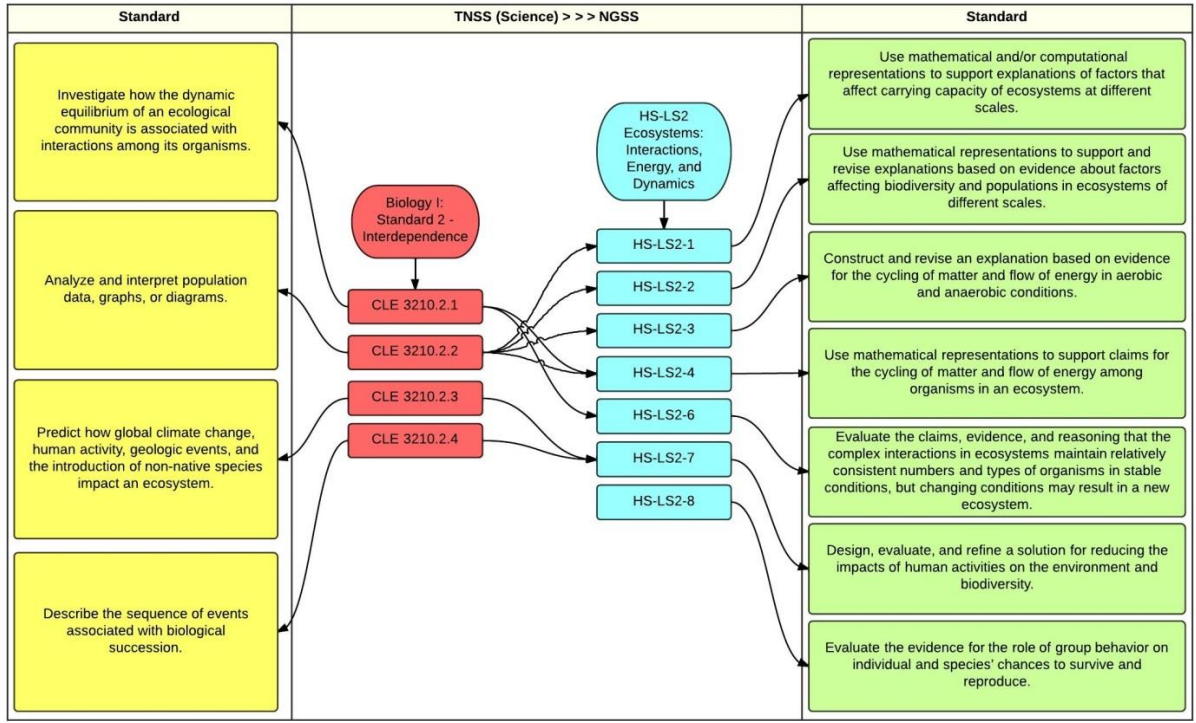

**Figure 1.3:** High School Biology Guidemap for Cellular and Molecular Biology (Part 3), Tennessee Department of Education Curriculum Standards in Science Education (left center) with explanations (far left) correspondence to Next Generation Science Standards (right center) with explanations (far right).

**Figure 2**

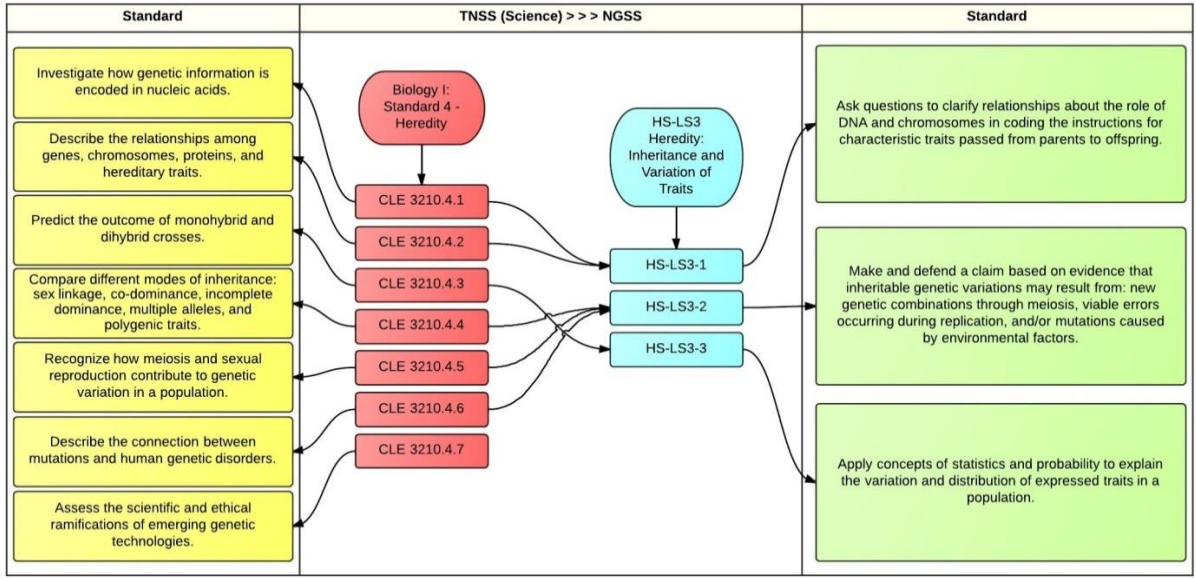

**Figure 2:** High School Biology Guidemap for Genetics and Inheritance, Tennessee Department of Education Curriculum Standards in Science Education (left center) with explanations (far left) correspondence to Next Generation Science Standards (right center) with explanations (far right).

**Figure 3**

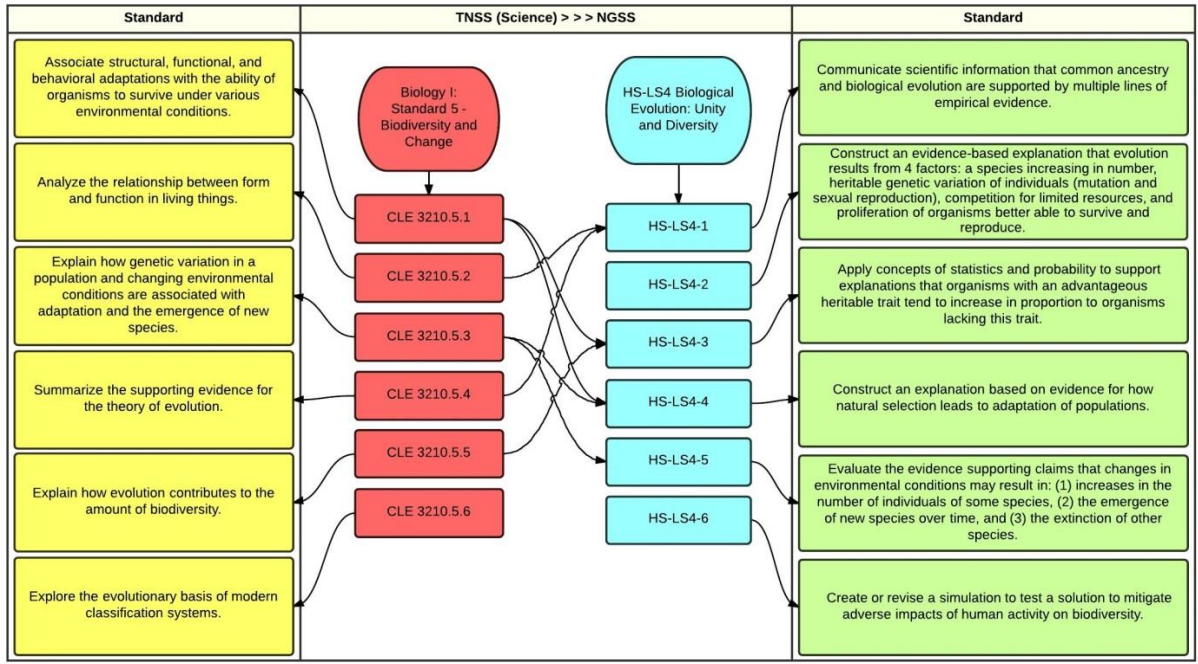

**Figure 3:** High School Biology Guidemap for Ecology, Evolution, and Biodiversity, Tennessee Department of Education Curriculum Standards in Science Education (left center) with explanations (far left) correspondence to Next Generation Science Standards (right center) with explanations (far right).

*TNCSSE-NGSS Guidemaps*

A series of posters showcase correlations between the current Tennessee Department of Education Curriculum Standards for Science Education (TNCSSE) and Next Generation Science Standards (NGSS). This product is an outcome of *Science First!*, a grant funded through the National Science Foundation Division of Graduate Education; (Grant Number DGE-0742364; P.I. Dr. Gordon Anderson). This NSF GK-12 Graduate Fellowship Program is supported by East Tennessee State University in partnership with North Side School of Math, Science, and Technology, a high need and racially/ethnically diverse school. *Access website:* http://www.etsu.edu/cas/gk/.

The sequence of 18 posters serves as a series of guidemaps between the TNCSSE for grade levels kindergarten through high school and the corresponding NGSS (See http://www.netstemhub.com). The TNCSSE-NGSS Guidemaps are designed with ease of use in mind. Care was taken to ensure that the guidemaps are a more effective way to correlate standards than perusing through either state or NGSS's websites. Educators can use the guidemaps as a way to adapt their current and past lesson plans to NGSS standards by easily visualizing inclusions and exclusions between the two sets.

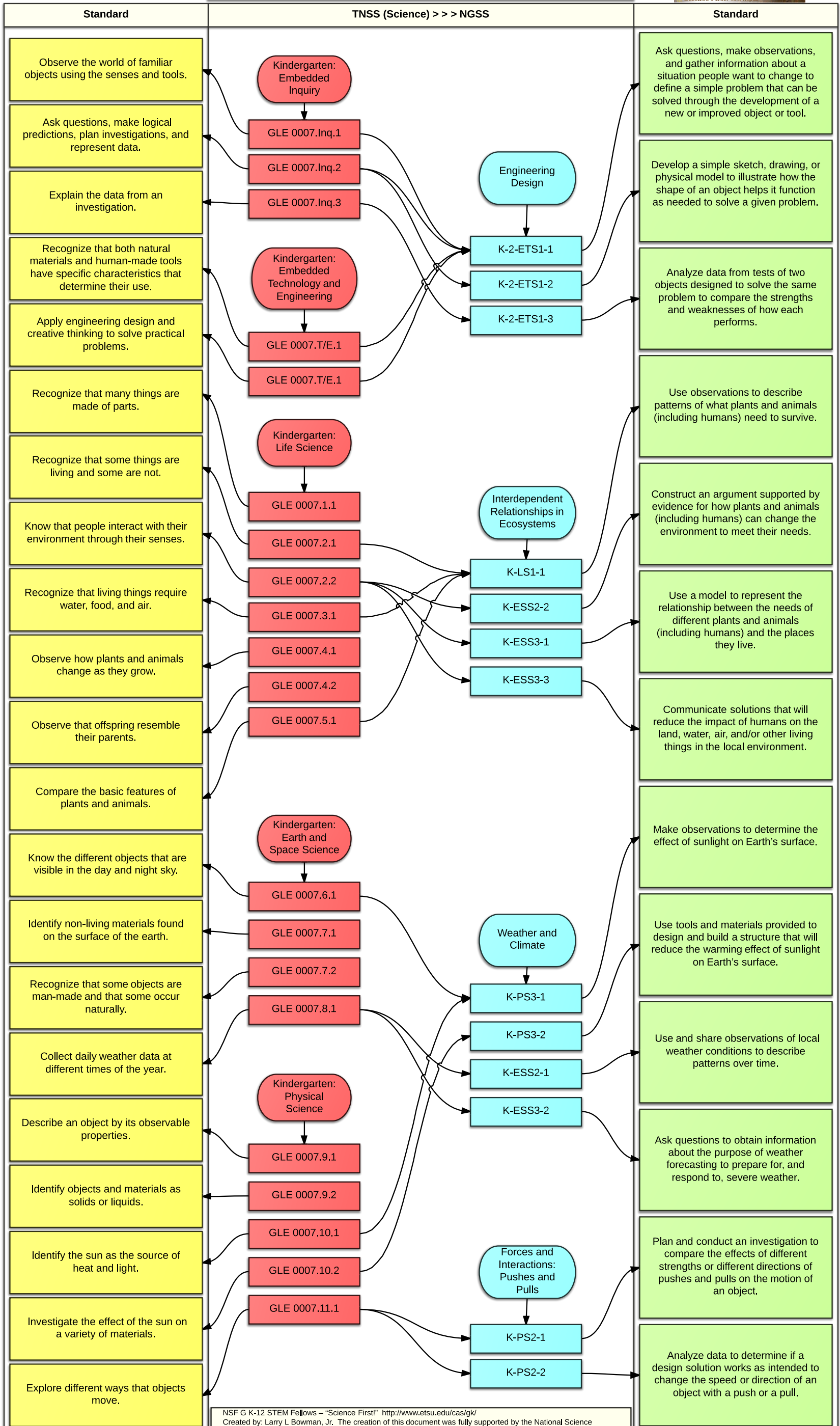

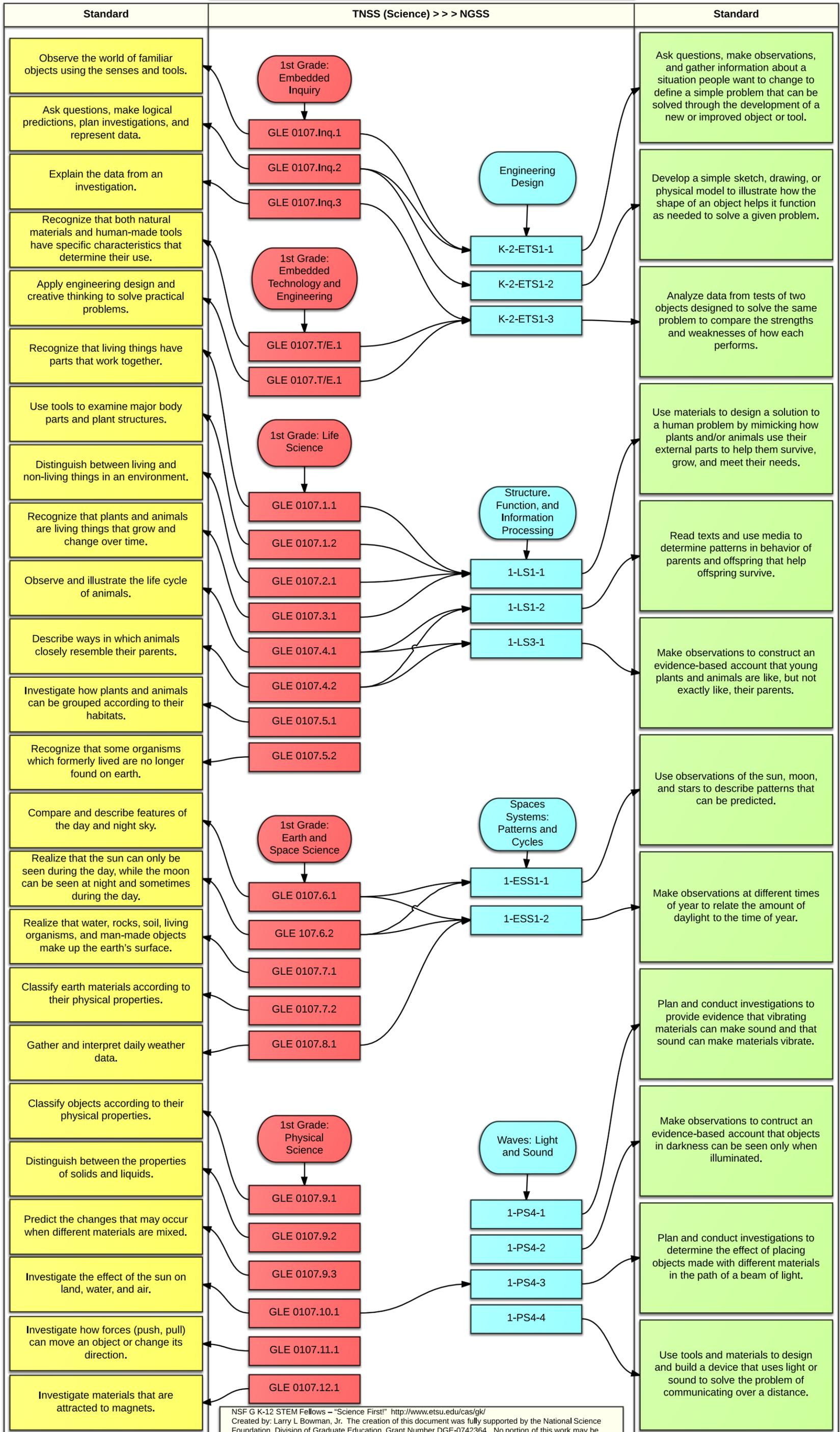

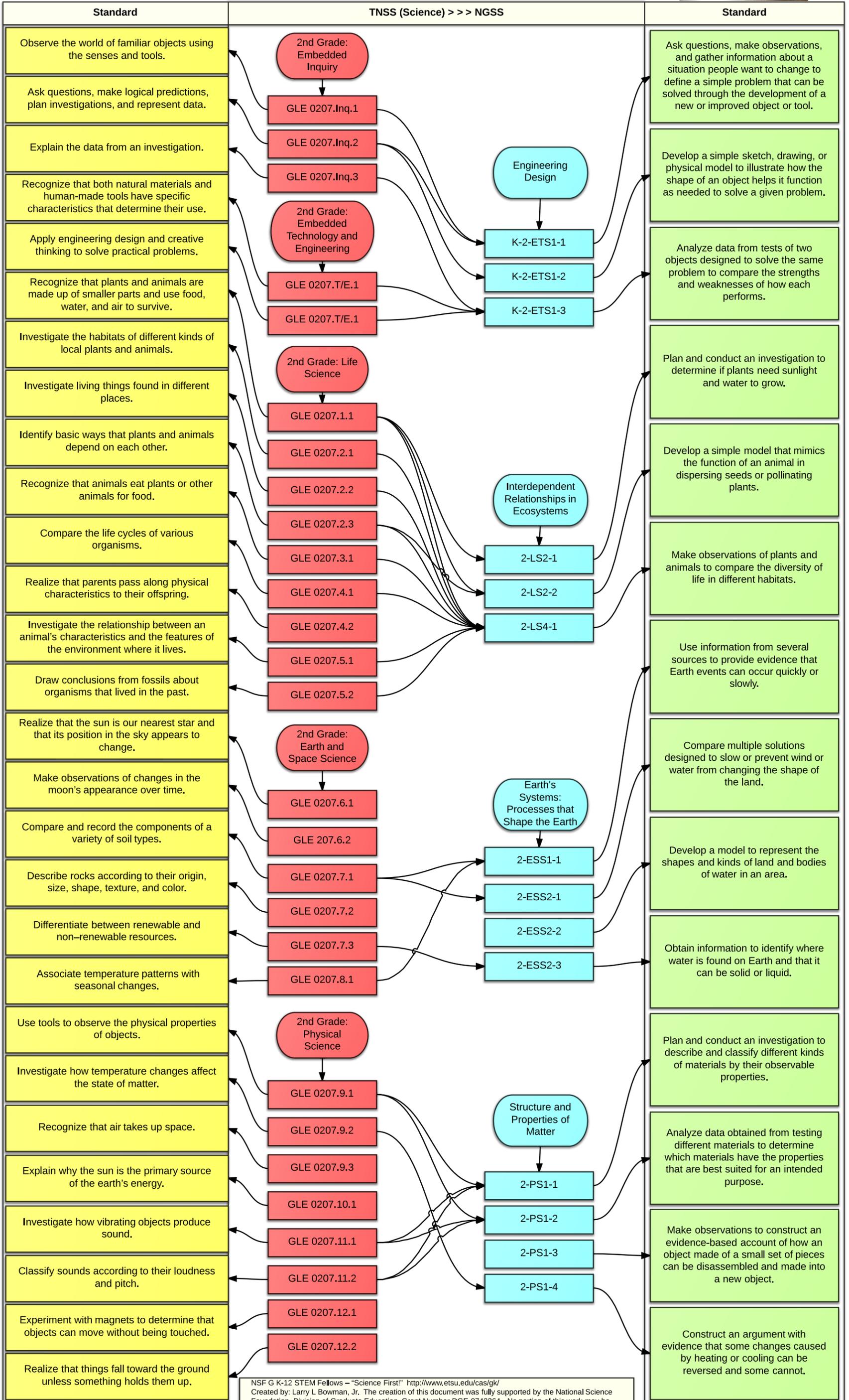

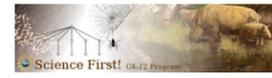

# 3rd Grade Map (Page 1)

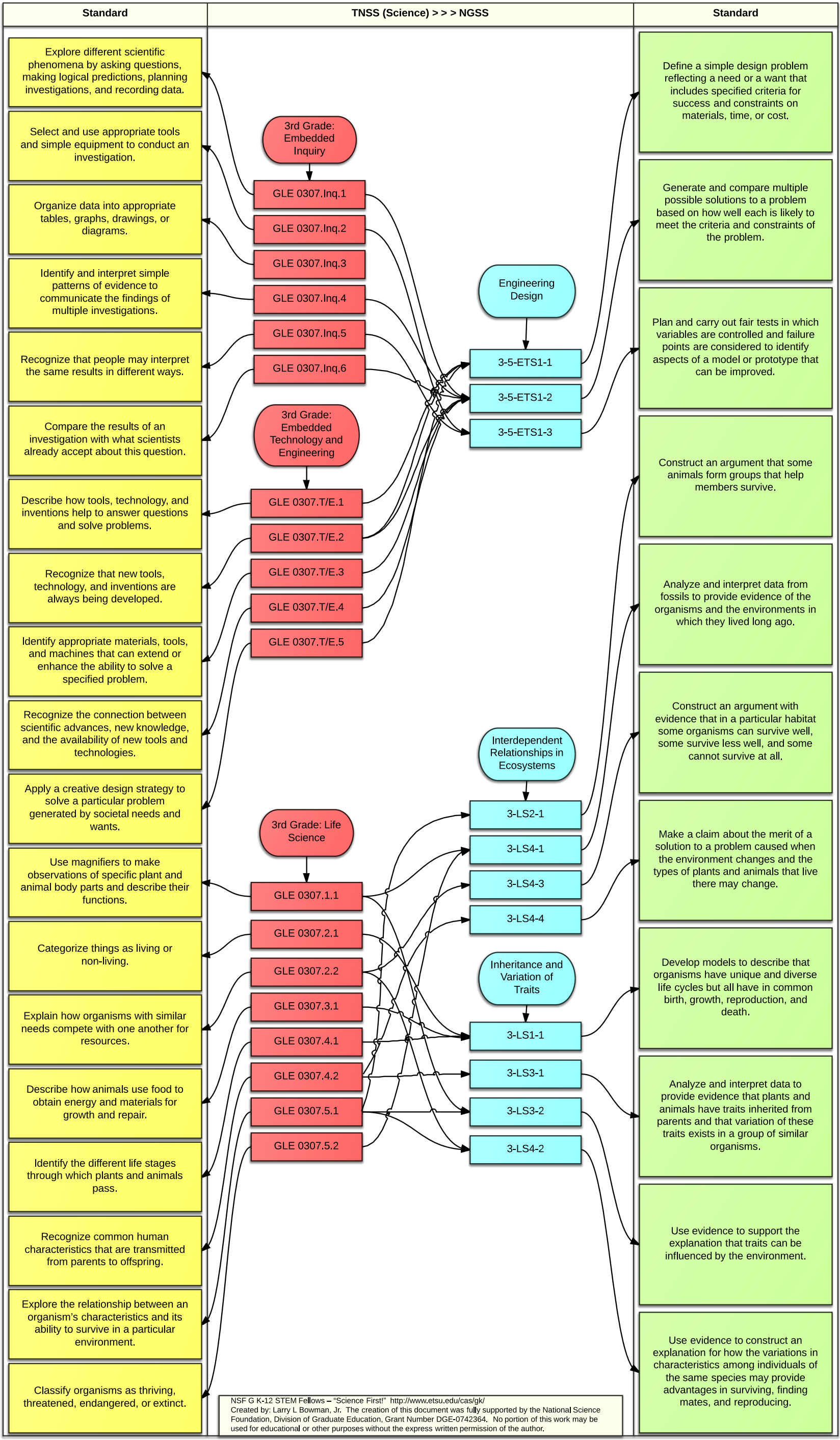

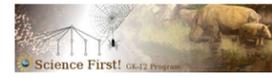

# 3rd Grade Map (Page 2)

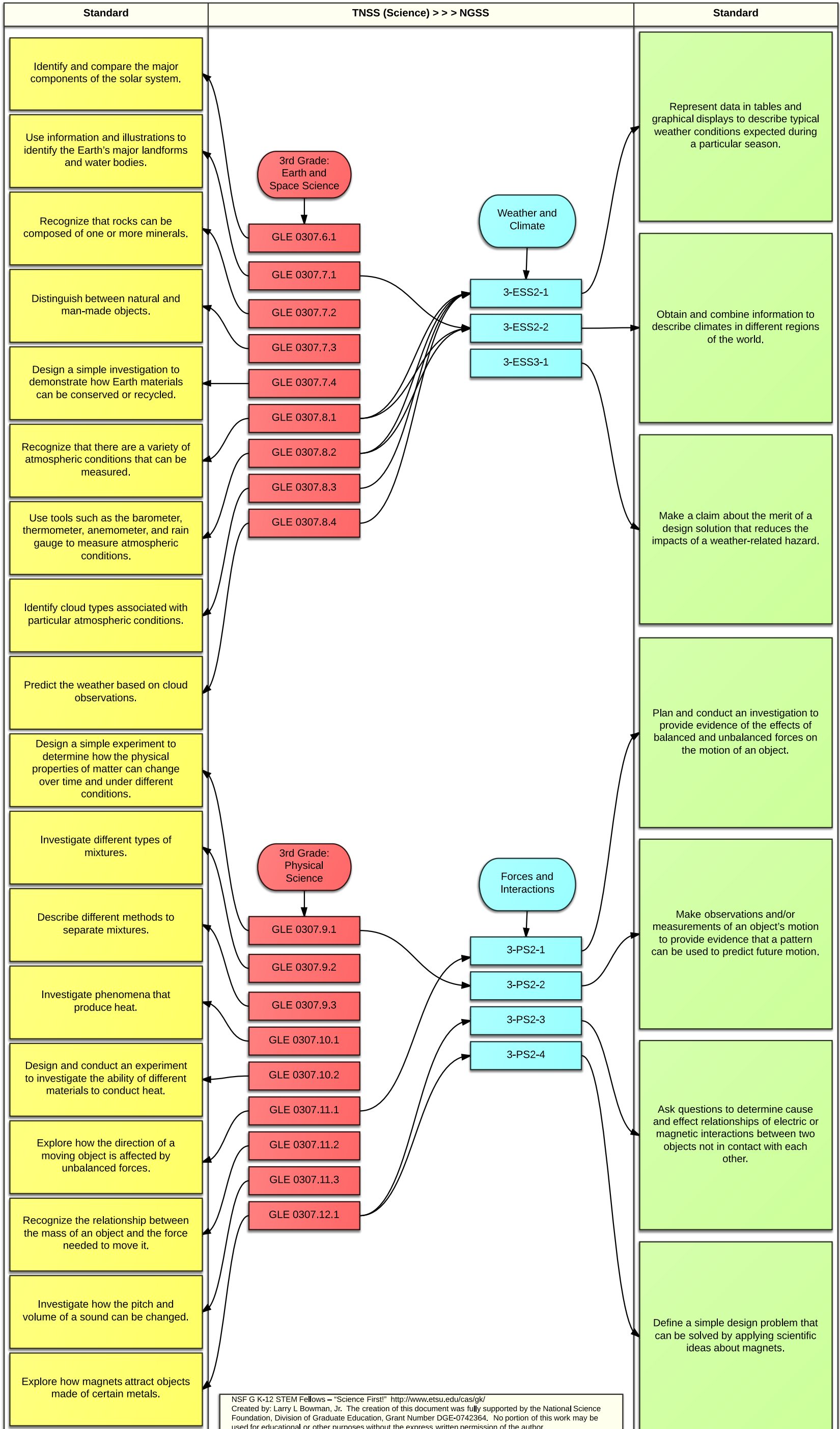

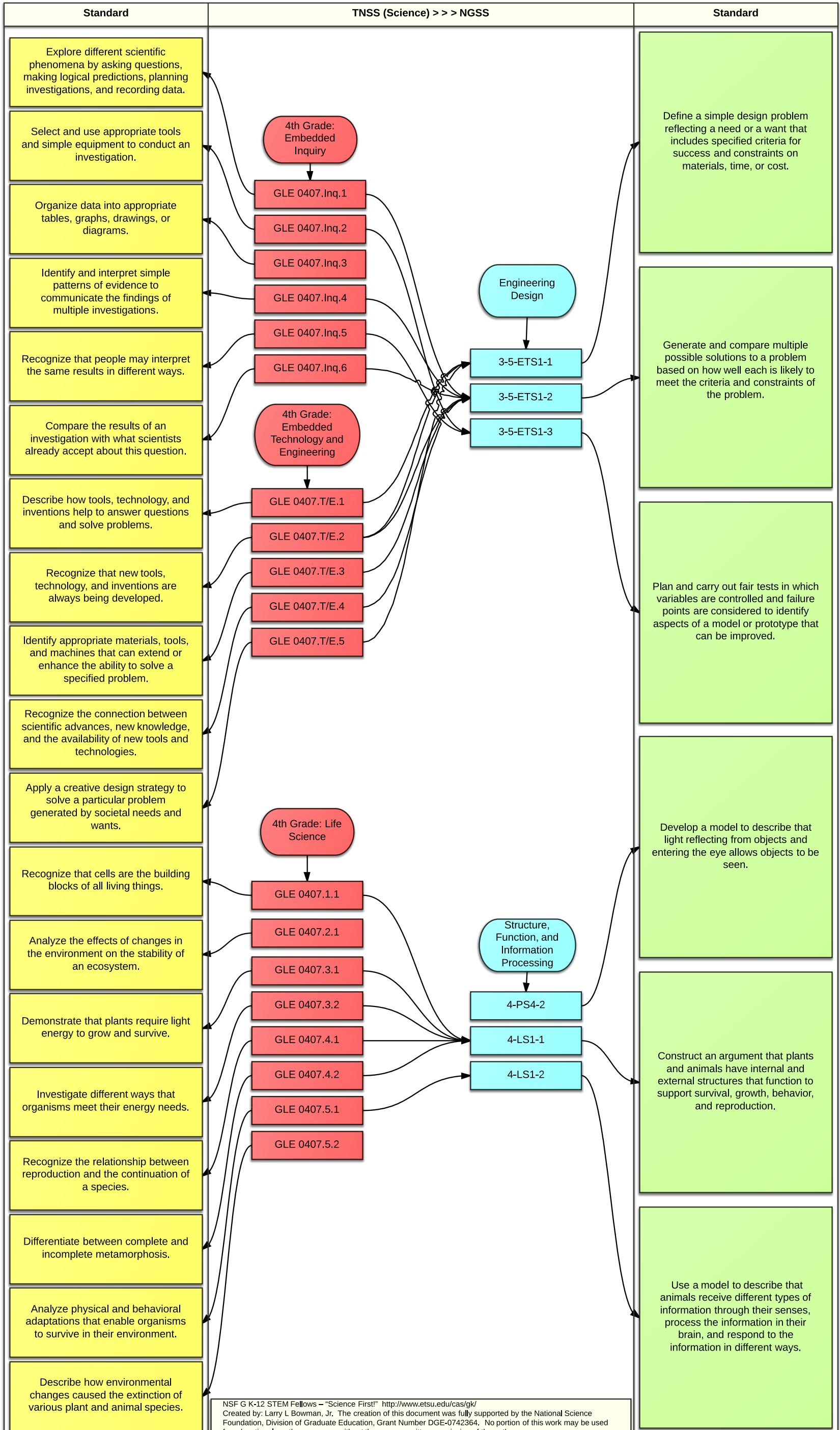

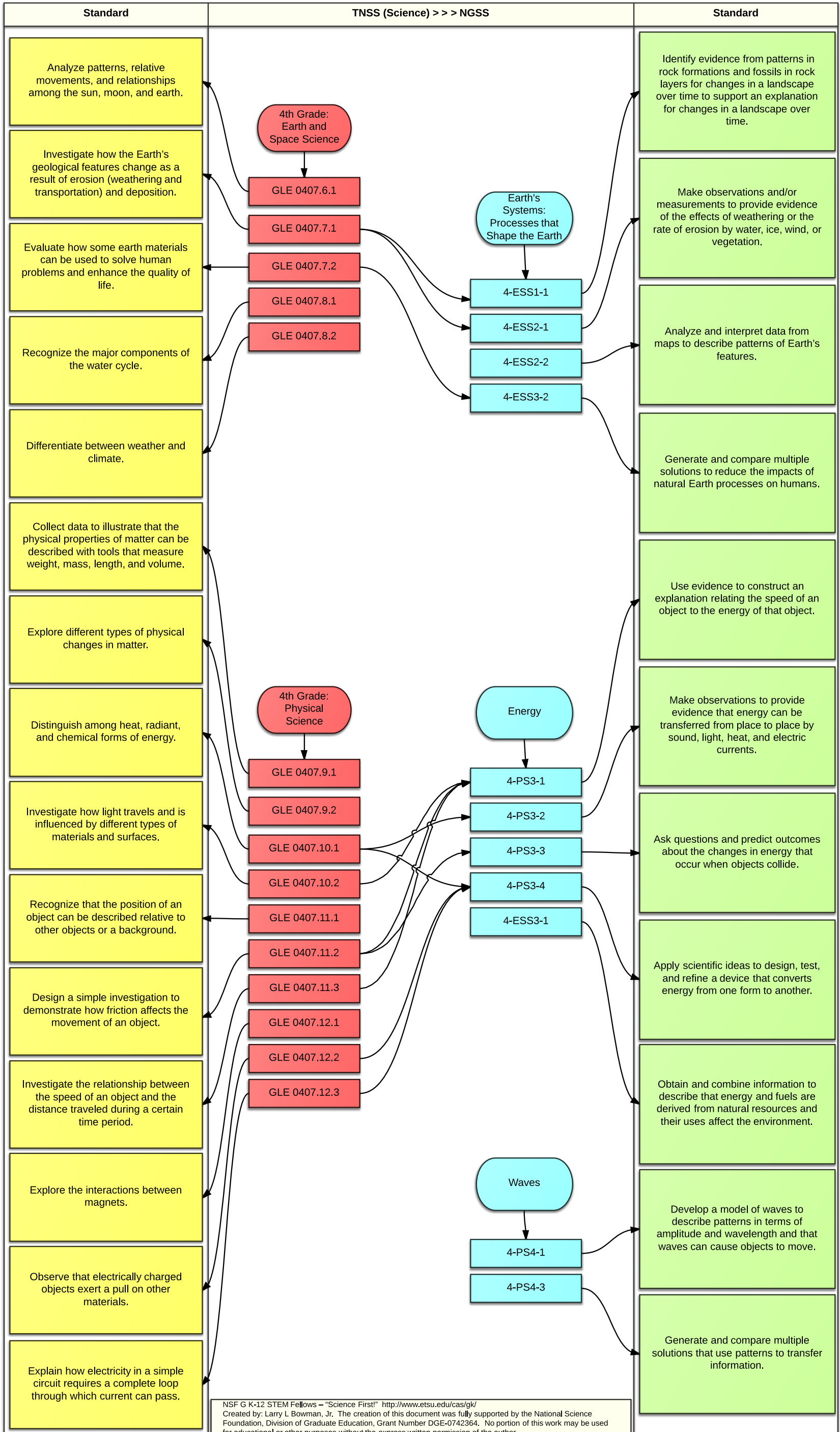

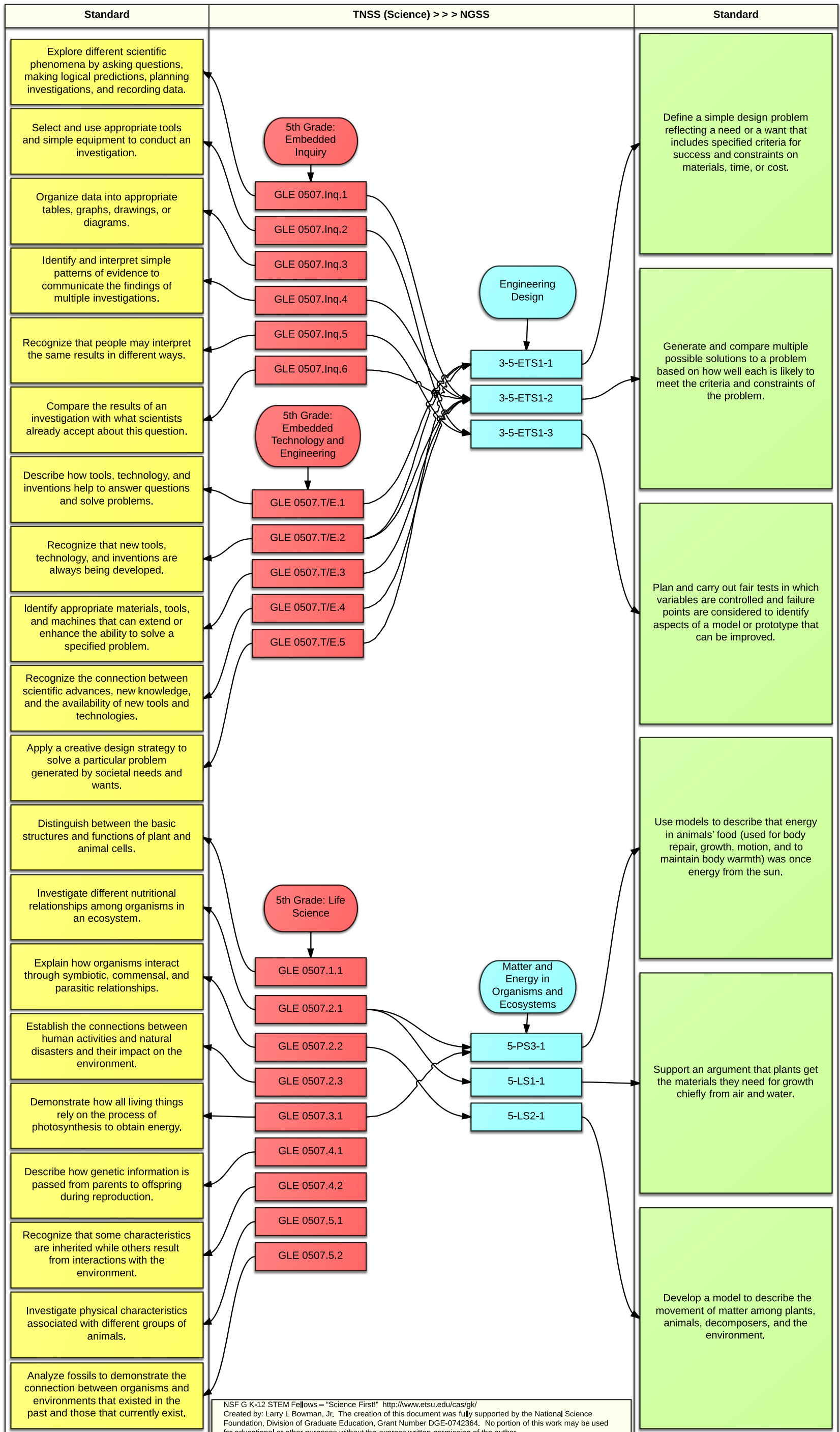

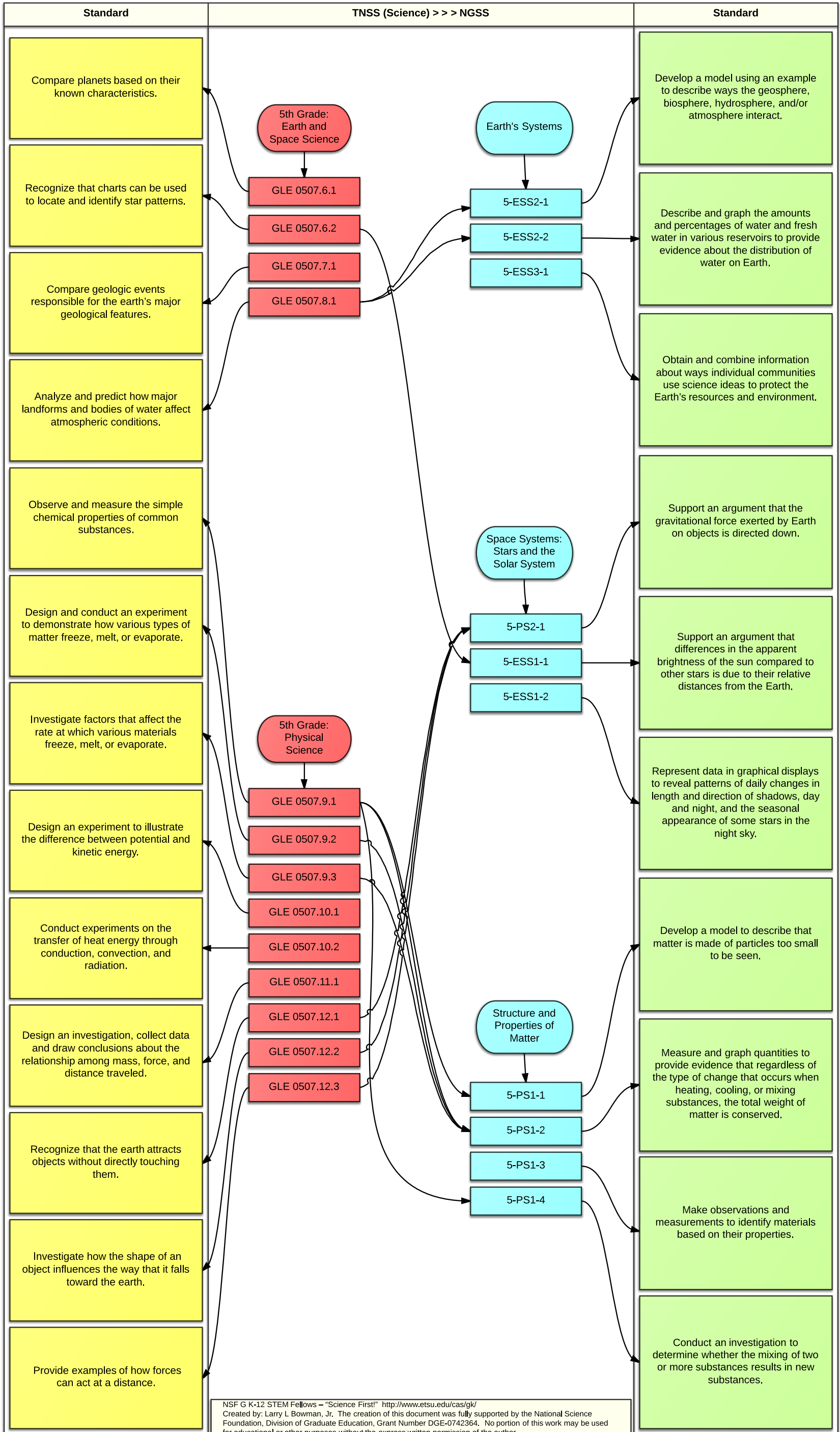

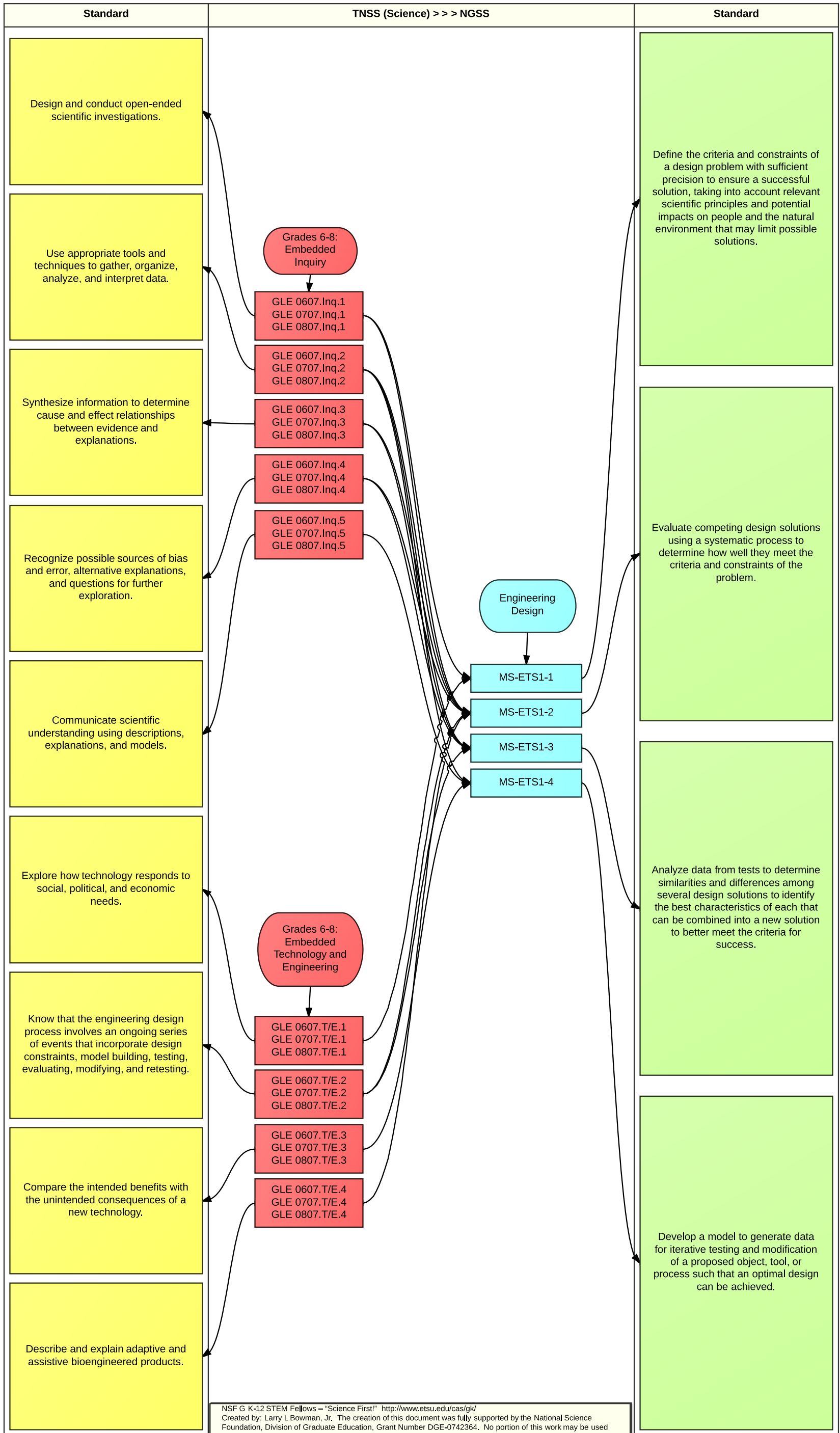

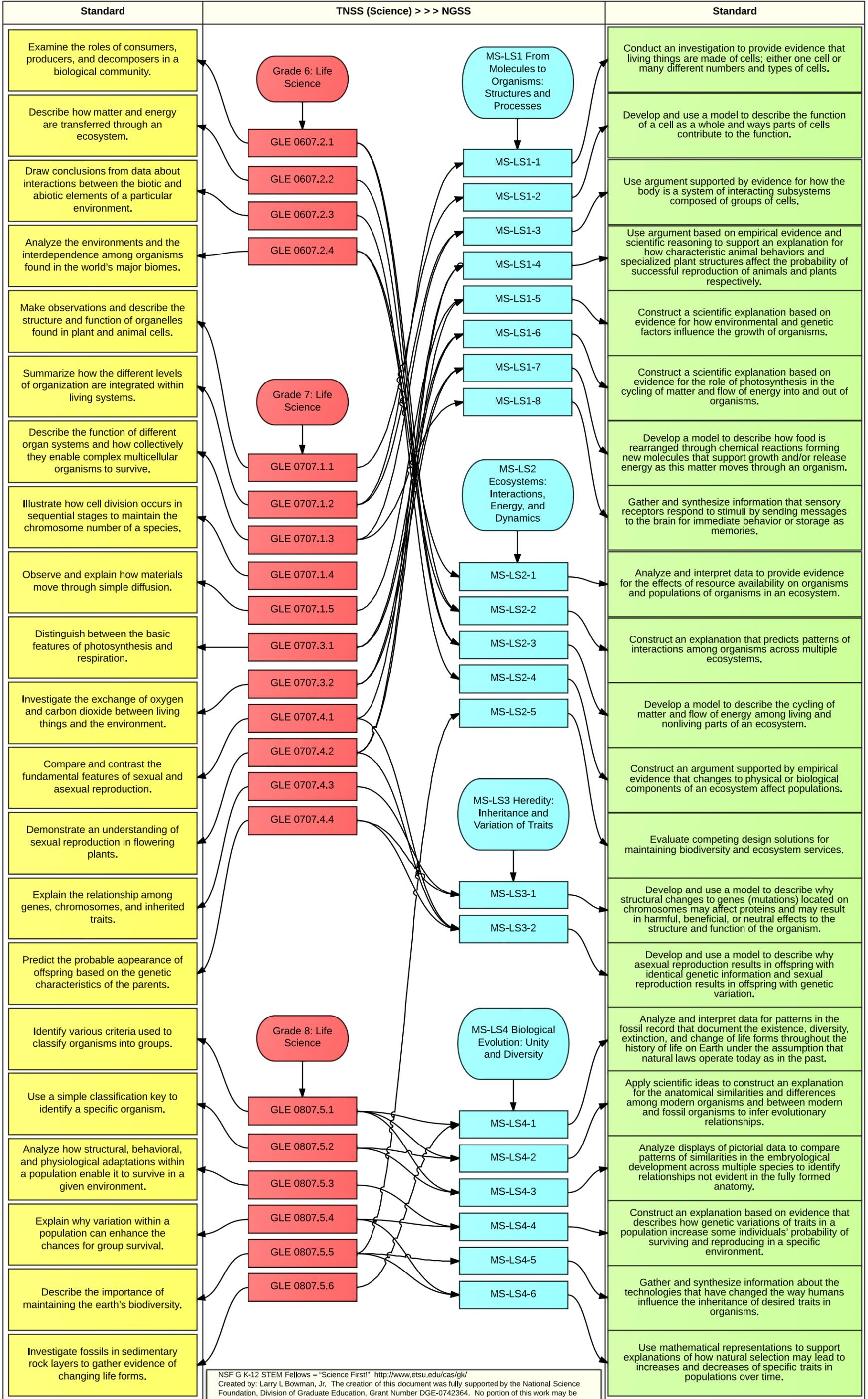

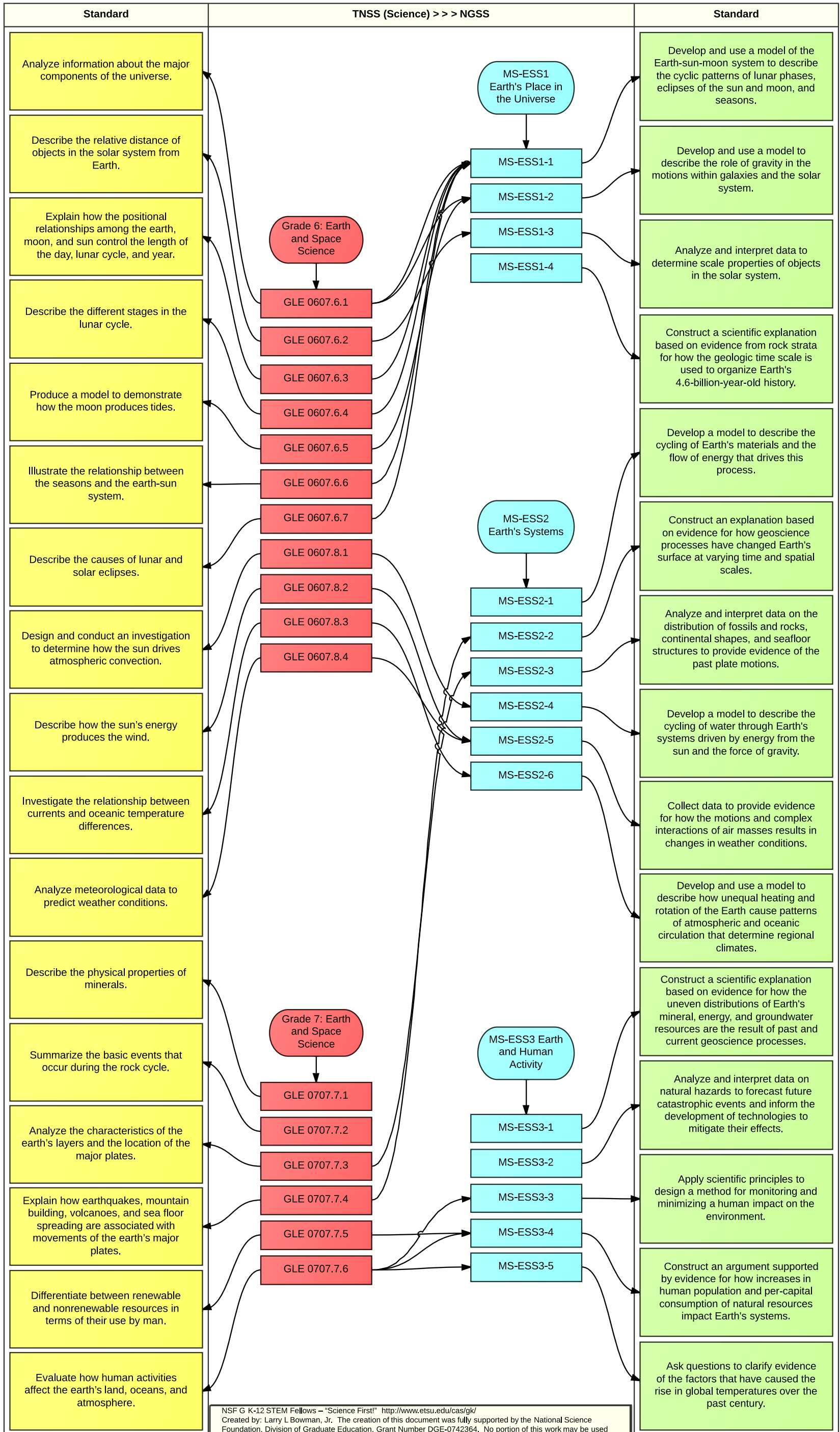

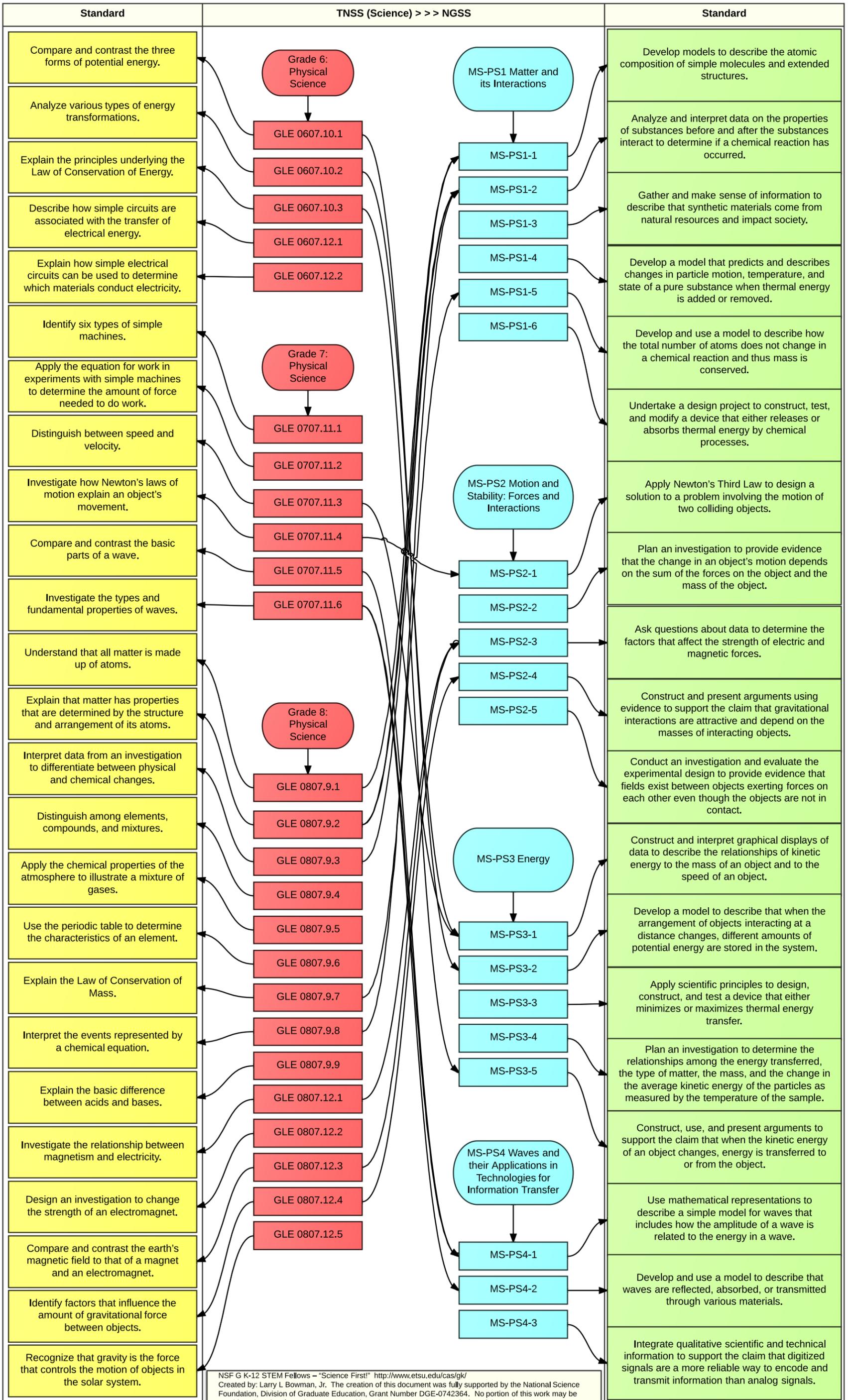

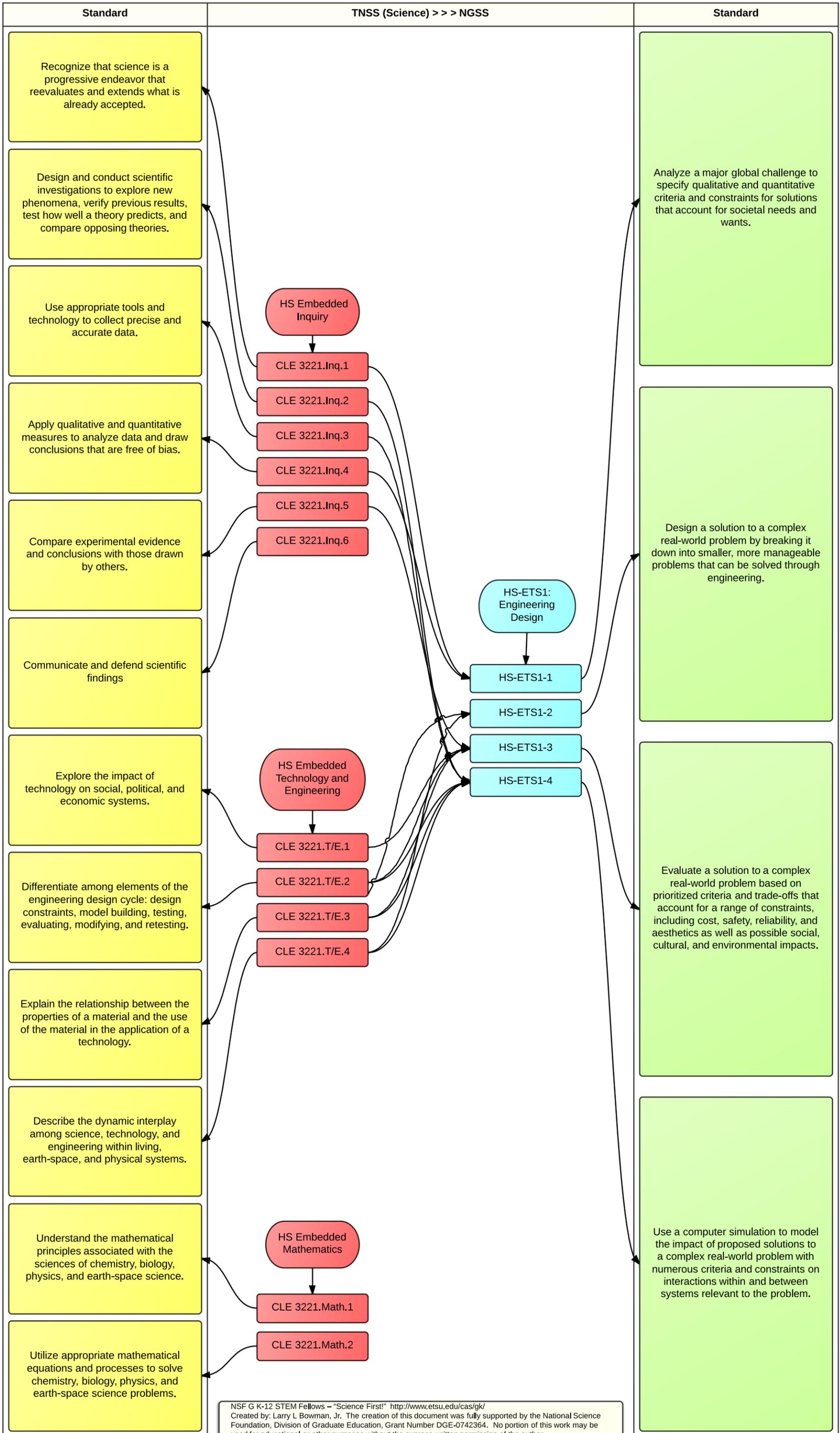

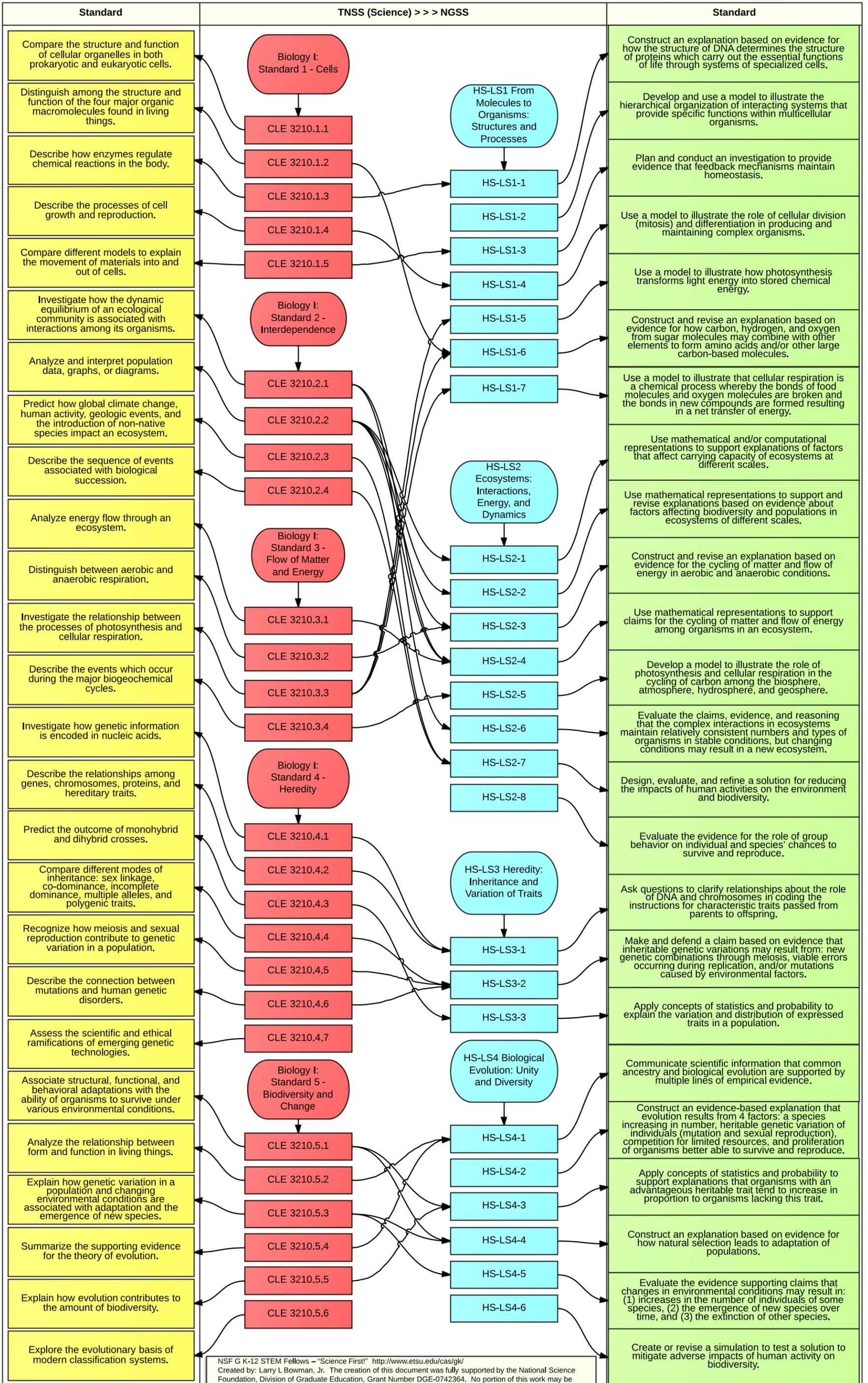

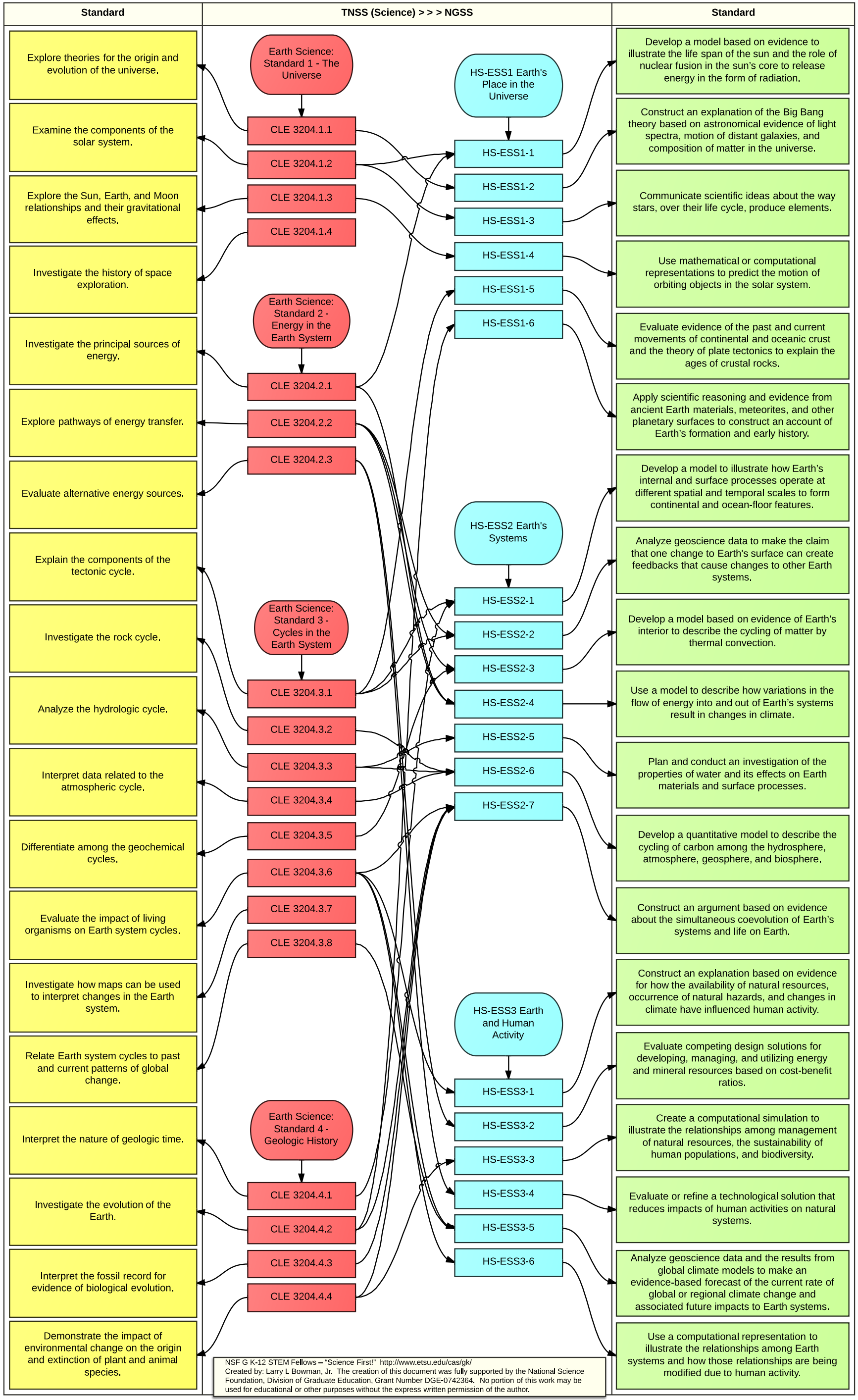

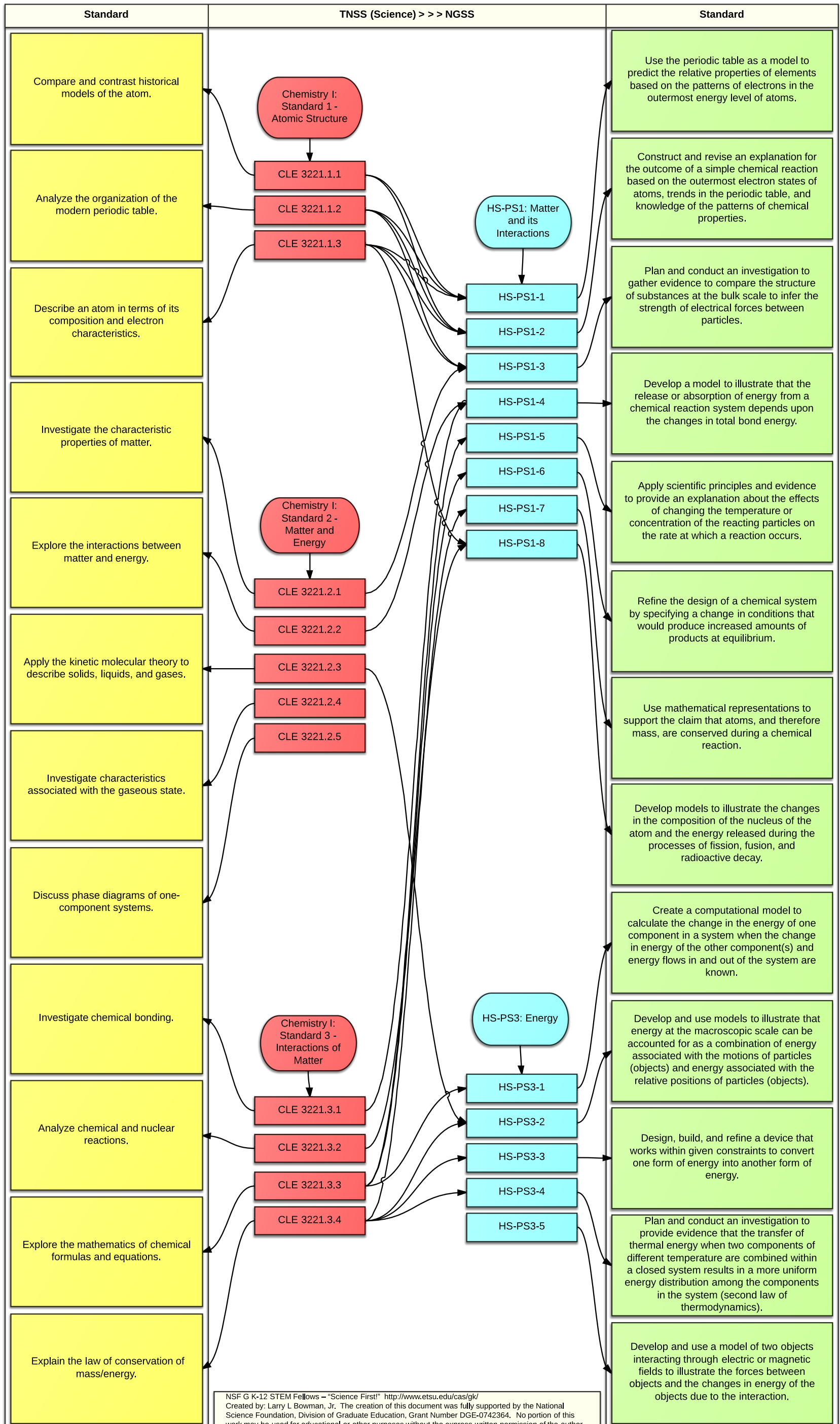

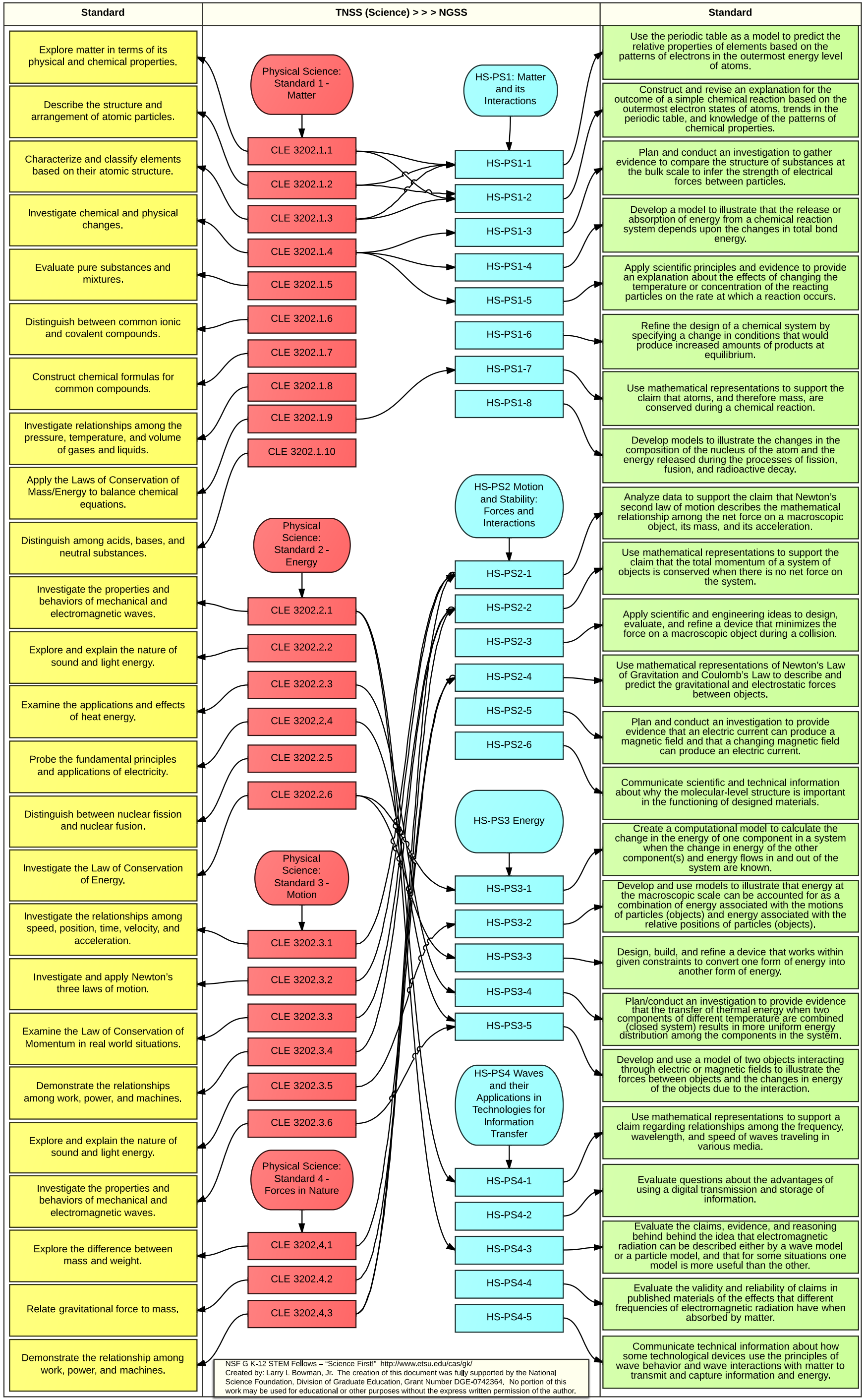